\documentclass[fleqn,usenatbib]{mnras}

\usepackage{mathptmx}

\usepackage[T1]{fontenc}

\DeclareRobustCommand{\VAN}[3]{#2}
\let\VANthebibliography\thebibliography
\def\thebibliography{\DeclareRobustCommand{\VAN}[3]{##3}\VANthebibliography}

\usepackage{graphicx}	
\usepackage{amsmath}	
\usepackage{amssymb}	
\usepackage{xcolor}
\usepackage[normalem]{ulem}
\usepackage{academicons}
\usepackage{orcidlink}

\newcommand{\jed}[1]{{\color{violet}#1}}
\newcommand{\msun}{ M$_{\odot}$}

\title[GCs of UDGs]{Imposters among us: globular cluster kinematics and the halo mass of ultra-diffuse galaxies in clusters}

\author[J. E. Doppel et al.]{
Jessica E. Doppel,$^{\orcidlink{0000-0001-5354-4229}1,2,3}$\thanks{E-mail: jessicadoppel@gmail.com}
Laura V. Sales,$^{\orcidlink{0000-0002-3790-720X}1}$
Jos\'e A. Benavides,$^{\orcidlink{0000-0003-1896-0424}1}$
Elisa Toloba,$^{\orcidlink{0000-0001-6443-5570}4}$
Eric W. Peng,$^{\orcidlink{0000-0002-2073-2781}5}$
\newauthor Dylan Nelson$^{\orcidlink{0000-0001-8421-5890}6}$,
and Julio F. Navarro$^{\orcidlink{0000-0003-3862-5076}7}$
\\
$^{1}$University of California, Riverside, 900 University Ave, Riverside, CA 92521, USA\\
$^{2}$Institute for Computational Cosmology, Department of Physics, University of Durham, South Road, Durham DH1 3LE, UK\\
$^{3}$Centre for Extragalactic Astronomy, Department of Physics, University of Durham, South Road, Durham DH1 3LE, UK\\
$^{4}$Department of Physics, University of the Pacific, 3601 Pacific Avenue, Stockton, CA 95211, USA\\
$^{5}$NSF’s National Optical-Infrared Astronomy Research Laboratory, 950 North Cherry Avenue, Tucson, AZ, 85719, USA\\
Riverside, CA 92521, USA\\
$^{6}$Institut fur theoretische Astrophysik, Zentrum fur Astronomie, Universitat at Heidelberg, D-69120 Heidelberg, Baden-Wurttemburg, Germany\\
$^{7}$Department of Physics and Astronomy, University of Victoria, Victoria, BC, Canada V8P 5C2
}

\date{Accepted XXX. Received YYY; in original form ZZZ}

\pubyear{2015}

\begin{document}
\label{firstpage}
\pagerange{\pageref{firstpage}--\pageref{lastpage}}
\maketitle

\begin{abstract}
 The velocity dispersion of globular clusters (GCs) around ultra-diffuse galaxies (UDGs) in the Virgo cluster spans a wide range, including cases where GC kinematics suggest halos as massive as (or even more massive than) that of the Milky Way around these faint dwarfs. We analyze the catalogs of GCs derived in post-processing from the TNG50 cosmological simulation to study the GC system kinematics and abundance of simulated UDGs in galaxy groups and clusters. UDGs in this simulation reside exclusively in dwarf-mass halos with $M_{200} \sim 10^{11}$\msun. When considering only GCs gravitationally bound to simulated UDGs, we find GCs properties that overlap well with several observational measurements for UDGs. In particular, no bias towards overly massive halos is inferred from the study of bound GCs, confirming that GCs are good tracers of UDG halo mass. However, we find that contamination by intra-cluster GCs may, in some cases, substantially increase velocity dispersion estimates when performing projected mock observations of our sample. We caution that targets with less than $10$ GC tracers are particularly prone to severe uncertainties.Measuring the stellar kinematics of the host galaxy should help confirm the unusually massive halos suggested by GC kinematics around some UDGs.
 
\end{abstract}




\section{Introduction}

Ultra-diffuse galaxies (UDGs), galaxies of extremely low surface brightness for their stellar mass, are enigmatic systems whose origin remains unclear. While the presence of such objects has been known for several decades \citep[see e.g. ][]{Reaves1983, Binggeli1985, Impey1988, Bothun1991, Dalcanton1997}, they have only recently entered the realm of systematic study with the observation of many UDGs in the Coma cluster \citep[see][]{Abraham2014, vanDokkum2015a, vanDokkum2015b}. 
 
UDGs were thought to reside primarily in the environments of galaxy clusters \citep[see][]{vanDokkum2015a, vanDokkum2015b, Koda2015, Mihos2015, Peng2016, Yagi2016, Gannon2022}, but 
they have since been observed in a much wider range of environments \citep{vanderBurg2017, Lee2017, Lee2020, Marleau2021, LaMarca2022, Venhola2022}, including in the field \citep{MartinexDelgado2016, Roman2017, Leisman2017, MartinNavarro2019, Rong2020b}. While many are observed to be devoid of gas \citep{MartinexDelgado2016, Papastergis2017, Roman2019, Junais2021}, more recent observations find gas-rich UDGs 
\citep[e.g., ][]{Leisman2017, ManceraPina2020, Jones2023}.
In addition to spanning a wide range of gas fraction and environments, UDGs also broadly span nucleation fraction \citep{Lim2020}.

Given the apparent diversity of UDGs, it has proven particularly difficult to pinpoint a unique formation path that may explain their origin. Several theoretical and numerical studies have pointed to differences between the dark matter halos that host UDGs and normal dwarfs, suggesting the possibility that UDGs may reside in dark matter halos with higher-than-average spin \citep{Amorisco2016, Rong2017, ManceraPina2020, Kong2022, Benavides2023}. Other studies present more baryon-focused formation scenarios. Star formation and feedback processes associated with starburst-driven outflows have the potential to leave the stellar component of galaxies rather extended \citep[e.g., ][]{DiCintio2017, Chan2018}, although even galaxies passively forming stars have been shown to form UDGs \citep{Tremmel2020}. To add an additional complication in the search for UDG formation, environmental effects, such as tidal heating \citep{Carleton2019} and tidal stripping \citep{Maccio2021, doppel2021, Moreno2022}, have also been argued to give rise to UDG-like galaxies. Moreover, combinations of the aforementioned scenarios are also possible \citep{Jiang2019, Sales2020}, thus an obvious UDG-formation route has yet to emerge.

Constraining the dark matter content of UDGs provides an additional dimension to understanding the origin of UDGs. For example, UDGs with little to no dark matter could suggest a primary formation mechanism of tidal stripping or other processes that preferentially removes dark matter) as a main driver \citep[see e.g.,][]{vanDokkum2018, vanDokkum2019, vanDokkum2022, TrujilloGomez2022}. At the other extreme, UDGs that inhabit overly-massive halos for their stellar mass could indicate that UDGs may originate as systems originally destined to become large, massive galaxies but where star formation was truncated early on \citep[see e.g.,][]{Forbes2020, vanDokkum2017, vanDokkum2015b, Toloba2023}. Between these two extremes, UDGs that reside in dark matter halos on par with other those of galaxies of similar stellar mass could suggest that UDGs are simply the tail of the surface brightness distribution of normal galaxies, and thus lack a distinct origin\citep[e.g.][]{Toloba2018, Lee2017, Lee2020, Saifollahi2021, Toloba2023}. Illuminating the dark matter content of UDGs is, therefore, a necessary component for pinpointing the---potential spectrum of---formation scenarios through which UDGs may arise and help to solidify their place in our understanding of dwarf galaxies.

Unfortunately, the dark matter content reported thus far for UDGs is as varied as their potential formation scenarios. Observations of luminous, kinematical tracers such as stars (e.g., DF44 \citep{vanDokkum2017} and DF4 \citep{Danieli2019} among others), globular clusters (GCs) \citep[see e.g.][]{vanDokkum2018, Toloba2018, vanDokkum2019}, and gas \citep{ManceraPina2020} suggest that the dark matter halos of UDGs span the entire range between lacking dark matter (such as DF2 and DF4 in NGC1052,)
to residing in halos with masses far exceeding those expected for their stellar masses \citep{Beasley2016, Janssens2022, Gannon2023, Toloba2023}, with others between these extremes \citep[see e.g.][]{Lee2017, Toloba2018, Lee2020, Saifollahi2021, Toloba2023}.

For UDGs for which kinematical tracers, such as stars and gas, are unavailable, 
globular clusters (GCs) offer an alternative measure of their halo masses due to their relative ease of observation over large distances and their rather extended spatial distributions. The numerous GCs often associated to UDGs have been intepreted to indicate that they reside in over-massive dark matter halos \citep{vanDokkum2015a, Peng2016, vanDokkum2017, Lim2018, Lim2020, Danieli2022, Janssens2022} if the power-law relation between GC mass and halo mass \citep[see e.g.][]{Peng2008, Harris2015} holds for UDGs. However, recent observations from the Coma cluster suggest that, by GC counts, there appears to be two types of UDGs: those that reside in apparently over-massive halos for their stellar mass, and those that appear to reside in halos of more typical in mass for dwarf galaxies \citep{Lim2018, Muller2021, Forbes2020, Jones2023}. 
A further characterization, as well as the theoretical context, of the observations of the GC systems of UDGs will further help to disentangle the dark matter component of UDGs.

With the high resolution of the TNG50 simulation of the IllustrisTNG suite, it is possible to morphologically define a set of simulated UDGs with similar structural parameters to observed UDGs \citep{Benavides2023}. Coupled with the recent addition of a catalog of GCs added to the simulation \citep{Doppel2023}, we can investigate UDGs in conjunction with their GC systems across a variety of environments, ranging from those comparable with massive elliptical systems to those comparable with the mass of the Fornax and Virgo clusters. We can thus make a realistic comparison with the observations of the GC systems of UDGs in these types of environments to provide possible interpretations for these observations and their implications for the dark matter content of UDGs.

In Section \ref{sec:methods}, we briefly discuss the details of TNG50 as well as the tagging model used to produce its GC catalog. In Section \ref{sec:gc_abundance}, we discuss how the modeled GC abundances and kinematics compare to observations as well as what, if any, effect environment has on UDGs and their GC systems. In Section \ref{sec:mock}, we compare mock observations of the GCs and UDGs in TNG50 to observed UDGs, and we use those mock observations to understand the inferred dark matter content of UDGs, both in the presence of contamination in their assigned GC systems as well as other complicating factors. Finally, in Section \ref{sec:summary}, we provide a short discussion and summary of our results.

\section{Methods}
\label{sec:methods}

\subsection{Simulation}
For this study, we use the highest resolution run of the cosmological hydrodynamical TNG50 \citep[][]{pillepich2019, nelson2019} simulation---which is part of the larger IllustrisTNG project \citep{naiman2018, pillepich2018, nelson2018, springel2018, marinacci2018, nelson2019dr}. TNG50 features a box size of 51.7 Mpc on each side with $2160^3$ gas cells and dark matter particles evolved assuming a flat, $\Lambda$CDM cosmology consistent with parameters from \citet{planckcollab2016}. This configuration results in a mass resolution of, on average, $8.4 \times 10^4 \ M_{\odot}$ for its baryonic component and $5.4 \times 10^5 \ M_{\odot}$ for dark matter particles. The gravitational softening length is $288$ pc at $z = 0$ for collisionless components.

The baryonic treatment in TNG50 is introduced in detail in \citep{weinberger2017, pillepich2018b}. Briefly, it includes star formation in the dense interstellar medium (ISM), stellar evolution, including chemical enrichment from stars and supernovae; primordial cooling, metal line cooling, and heating, via background radiation, of gas; additionally, the seeding and growth of supermassive black holes, low and high accretion AGN feedback, galactic winds, and magnetic fields \citep{weinberger2017, pillepich2018b}. 

\subsubsection{Sample Selection}

Halos and subhalos within the TNG50 simulation are identified using the Friends-of-Friends \citep[FOF,][]{davis1985} and SubFind \citep{springel2001, dolag2009} respectively. Using these catalogs, we select 39 halos with virial masses between $M_{200} = [5\times 10^{12}, 2\times 10^{14}$]\msun (where ``virial'' in this study refers to quantities associated to a sphere enclosing 200 times the critical density of the universe).
The mass resolution of TNG50 allows us to resolve galaxies with a stellar component of $M_* \sim 5\times 10^{6}$\msun, which therefore contain at least 60 stellar particles. A stricter resolution threshhold is considered for this study: we consider only UDGs in the stellar mass range $M_* = [10^{7.5}, 10^{9}]$\msun---which are resolved with a minimum of $\sim 375$ stellar particles---that also reside in galaxy groups and clusters. The evolution of these objects are followed using the SubLink merger trees \citep{rodriguez-gomez2015}.

\begin{figure}
    \centering    
    \includegraphics[width = \columnwidth]{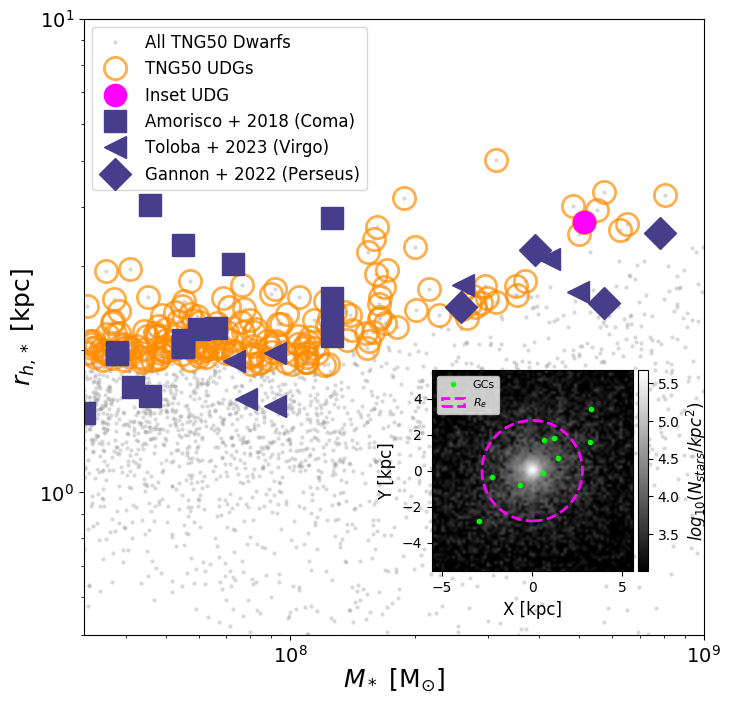}
    \caption{Stellar size ($r_{h,*}$) as a function of stellar mass ($M_{*}$) in TNG50 for all dwarf galaxies (gray dots), and for the UDG sample (unfilled orange circles). We show the same for UDGs in the Coma cluster \citep[purple squares, ][]{amorisco2018}, the Virgo cluster \citep[purple triangles][]{Toloba2023}, and the Perseus cluster \citep[purple diamonds, ][]{Gannon2022}. The size of observed UDGs has been multiplied by $4/3$ \citep[e.g. ][]{hernquist1990, Wolf2010, Somerville2018} to transform it into a 3D measurement (sec. \ref{sec:mock}). 
    Highlighted in pink is the size and mass of the example TNG50 UDG shown in projection in the inset panel, colored by stellar number density and overplotted with its 2D effective radius, $R_e$ (dotted pink circle), and GC system (lime green dots). We can see that, where data is available, there is good agreement between the sizes of the observed satellite UDGs in galaxy clusters and the sample of satellite UDGs in TNG50.} 
    \label{fig:sample_UDGs}
\end{figure}

\begin{figure*}
    \centering
    \includegraphics[width = 0.992\columnwidth]{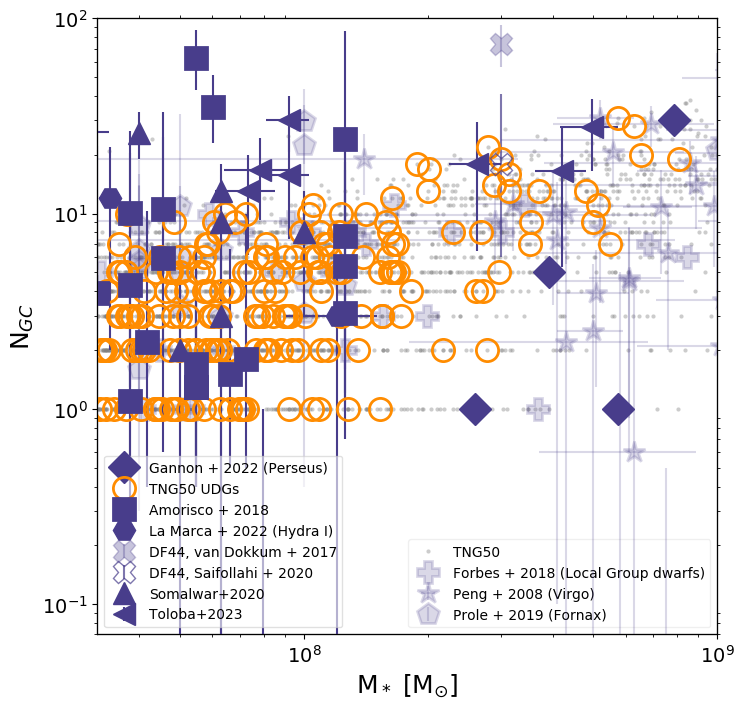}
    \includegraphics[width = \columnwidth]{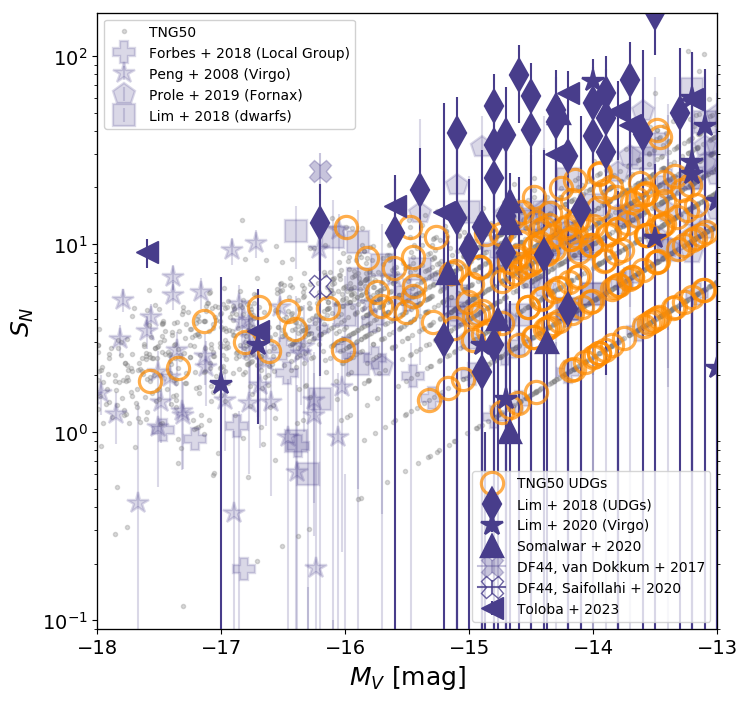}
    \caption{\textit{Left}: Number of GCs (N$_{GC}$) as a function of host galaxy stellar mass. All simulated TNG50 satellite dwarf galaxies are shown in translucent gray points, with UDGs highlighted by unfilled orange circles. Observations of GC numbers for normal dwarf galaxies are shown in purple, translucent shapes, and those for UDGs in purple, filled shapes. We can see that while there is a large amount of scatter in the predicted GC numbers for the UDGs of TNG50, the scatter is not as large as what is seen in observed UDGs, particularly those of the Coma Cluster (filled squares). We can see that despite the wide scatter, simulated and observational data follows (on average) similar trends. \textit{Right:} the specific frequency of GCs ($S_N$) as a function of host galaxy V-band absolute magnitude ($M_V$). Following \citet{Doppel2023}, we have applied a correction to the V-band magnitude to account for discrepancies between TNG50 and observations for high-mass galaxies. As in the left panel, all TNG50 dwarfs are shown as gray points, UDGs are highlighted by orange circles, observations of $S_N$ for normal dwarf galaxies are shown as translucent purple shapes, and observations of $S_N$ for UDGs are shown as filled, purple shapes. While the simulated UDGs seem to follow well the $S_N$ of observed normal dwarf galaxies and the bulk of observed UDGs, they are unable to reproduce the extreme $S_N$ for many UDGs in the coma cluster (filled purple diamonds).
    For both measures of GC abundance in the figures, there is significant overlap between what is predicted by TNG50 and what is observed for the bulk of UDGs; however, we do not predict the most extreme GC systems.}
    \label{fig:abundance}
\end{figure*}

\subsection{GC Catalog}
\label{ssec:gc_cat}
We use the GC catalog presented in \citet{Doppel2023}, which has been added in post-processing to the $39$ most massive galaxy groups and clusters in TNG50, spanning a virial mass range $M_{200} = [5\times 10^{12}, 2\times 10^{14}]$ \msun. 
GCs are tagged to all galaxies in the selected groups and clusters provided they satisfy a maximum stellar mass throughout their history of at least $5\times 10^6$ \msun~ and a minimum of $100$ dark matter particles (this latter condition is required to avoid spurious baryonic clumps). All galaxies are tagged at their infall time, which is here defined as the last time the galaxy is its own central. On average, this corresponds to the time at which a galaxy crosses the virial radius of its present day host halo, but it might be an earlier time if the galaxy joins a smaller halo or group before joining their final host system.

GC candidate particles are selected from the dark matter particles associated to the host galaxy at infall time. Following \citet{lokas2001}, we fit an NFW profile \citep{navarro1996}: 
    \begin{equation}
    \rho_{\rm NFW}(r) = \frac{\rho_{\rm NFW}^0}
    {(r/r_{\rm NFW})(1 + r/r_{\rm NFW})^2}
     \end{equation}
to the dark matter component of the galaxy. The scale radius $r_{NFW} = r_{max} / \alpha$, where $r_{max}$ is the radius of maximum circular velocity and $\alpha = 2.1623$ \citep{navarro1997}.

The GCs are assumed to follow a \citet{hernquist1990} profile:
\begin{equation}
    \rho_{\rm HQ}(r) = \frac{\rho_{\rm HQ}^0}{(r/r_{\rm HQ})(1 + r/r_{\rm HQ})^3}
    \end{equation}
which allows us to control the normalization and radial extension of the tagged GCs. We assign two populations of GCs: a red, metal-rich component of GCs that formed in-situ, and blue GCs, representative of older, more metal-poor GCs that were accreted into the galaxies. The red GCs are chosen to be more spatially concentrated than the blue GCs, with scale radii $r_{HQ} = 0.5 r_{NFW}$ and $3.0 r_{NFW}$ for red and blue GCs respectively, $\rho_{HQ}$ is chosen to maximize the number of GC candidates.

The GC candidates are then selected in relative energy using the distribution function \citep{bandt2008}:
\begin{equation}
    f_{i}(\epsilon) = \frac{1}{8\pi}\bigg[ \int_{0}^{\epsilon} \frac{\rm d^2\rho_i}{\rm d\psi^2} \frac{\rm d \psi}{\sqrt{\epsilon - \psi}} + \frac{1}{\sqrt{\epsilon}} \bigg(\frac{\rm d\rho_i}{\rm d\psi}\bigg) \bigg|_{\psi = 0}\bigg] ,
    \label{eqn:distfunc}
\end{equation}
where $\rho_i$ is the density profile of i = (dark matter, red GCs, and blue GCs), $\Psi$ is the relative gravitational potential, and $\epsilon$ is the relative energy. In equally spaced bins of relative energy, a fraction $f_{HQ,i}/f_{NFW}$, where i = red or blue GCs, of dark matter particles is selected. Inspired by constraints inferred for the Milky Way \citep{yahagi2005}, a cutoff radius of $r_h/3$, where $r_h$ is the total half-mass radius of the halo in question, for the GC candidate particles.

The selected GC candidate particles are assigned masses at infall such that by $z = 0$ those that still remain gravitationally associated to their host follow the $M_{GC} - M_{halo}$ relation from \citet{Harris2015}. To make this calibration, we assume that a power-law relation similar to the $M_{GC}-M_{halo}$ relation exists at infall such that:
\begin{equation}
    M_{\rm GC, inf} = \frac{1}{f_{\rm bound}} M_{\mathrm{GC}, z = 0} = a_{\rm inf} M_{\rm halo, inf}^{b_{\rm inf}} .
    \label{eq:calibration2}
\end{equation}
where $f_{bound}$ is the fraction of GCs that are still gravitationally bound to their host galaxy at $z = 0$. We find for red and blue GCs respectively, $a_{inf} = 2.6 \times 10^{-7}$ and $7.3 \times 10^{-5}$ and $b_{inf} = 1.14 \ \rm and \ 0.98$. 

Since the GC candidates are a much larger set of particles than the observed number of GCs, we subsample a realistic number of GCs from the candidates. This realistic population of GCs follows a Gaussian luminosity function using constraints from \citet{jordan2007}. Individual GC masses are obtained assuming a mass-to-light ratio of 1. GCs are randomly selected from the luminosity function until the total mass of GCs is within $7\times 10^3$ \msun~ (the assumed minimum mass of one GC) of the total calibrated infall mass. The realistic subsample of GCs is followed to $z = 0$ and constitutes the GCs we consider in this work.

\citep{Doppel2023} shows that this method reproduces the available observational constraints in number, specific frequency, and GC occupation fraction over a wide range of masses, including 
 dwarfs. In this paper we focus on the specific predictions of this GC catalog for the particular case of UDGs in galaxy groups and clusters. By design, our GC tagging method is able to capture the range in GC numbers and kinematics that is expected due solely to variations in the dark matter halos of UDGs at infall, being an excellent tool to guide the interpretation of current observations.

\subsection{Sample of UDGs in groups and clusters}
\label{ssec:selection} 
 The UDGs considered for this work are satellites of our selected galaxy groups and clusters and were first introduced in \citet{Benavides2023}. Simulated UDGs are selected to be in the stellar mass range $M_* = [10^{7.5}, 10^9]$\msun---to ensure that there are sufficient stellar particles to resolve the structure of the galaxy. 
 Inspired by the UDG classification process presented by \citet{Lim2020}, wherein UDGs are selected to be $2.5\sigma$ outliers in scaling relations between luminosity and surface brightness, mean effective surface brightness, and effective radius,
 UDGs are identified as the $5\%$ most extended outliers in the $M_*$-size relation. These UDGs are shown in Fig. \ref{fig:sample_UDGs}, which shows the relation between stellar halfmass radius, $r_{h,*}$ and stellar mass, $M_*$. These criteria result in UDGs that are roughly consistent with sizes of UDGs in \jed{the} Virgo \citep[purple triangles,][]{Toloba2023}, Coma \citep[purple squares][]{amorisco2018}, and Perseus \citep[purple diamonds][]{Gannon2022} clusters, low-density environments \citep{Roman2019, MartinNavarro2019, Rong2020a}, as well as other commonly assumed cut-offs to identify UDGs in observations ($R_{\rm e} \geq 1.5$ kpc and $\mu \gtrsim 24.5$ mag/arcsec$^2$ measured within the effective radius of stars \citep[e.g.][]{vanDokkum2015a}).

 As discussed in detail in \citet{Benavides2023}, the formation mechanism of UDGs in TNG50 suggests that they inhabit mainly high-spin dark matter halos, although a sub-dominant fraction ($\sim 10\%$) of satellite UDGs owe their extended sizes to tidal effects within their groups or clusters. Most importantly, all simulated UDGs in TNG50 formed within dark matter halos in the range $M_{200} \sim [10^{9.3} - 10^{11.2}]~$\msun~
 that are in agreement with expectations from their stellar content. In addition, satellite UDGs are found to be red and quiescent while field UDGs are gas-rich and star-forming, in good agreement with observational results \citep[e.g.][]{vanderBurg2016, Lee2020, FerreMateu2018, Leisman2017, ManceraPina2020, Jones2023}. Note that our simulations also predict a fraction of quiescent UDGs in the field as a result of backsplash orbits \citep{Benavides2021} that are not included in our sample as they, by definition, do not reside today within group or cluster halos.
 
 Satellite UDGs have typically undergone substantial tidal stripping of their dark matter halos (median mass loss $80\%$) but only moderate tidal stripping of their stellar component ($10\%$ mass loss from their peak stellar mass). A total of 195 UDGs are found associated to our simulated groups in TNG50 and are the core sample of the analysis in this paper. In addition, these groups and clusters have 2195 non-UDG dwarfs in the same mass range as our UDGs that might be included when necessary for helpful comparisons. \textit{This set of UDGs allows us the first opportunity to study the GC systems of UDGs that reside in realistic group and cluster environments.}


\section{GC abundance and kinematics in UDGs}
\label{sec:gc_abundance}

We show in Fig.~\ref{fig:abundance} the predicted GC number ($N_{GC}$, left panel) and GC specific frequency ($S_N$, right panel) for satellite dwarf galaxies in TNG50 compared to observational constraints. Specific frequency is defined as the number of GCs per unit luminosity normalized to a galaxy with V-band magnitude $M_v =-15$ as follows \citep{harris1981}:
\begin{equation}
    S_N = N_{GC} 10^{0.4(M_V + 15)}
\end{equation}
Overall, we find a good agreement between {\it all} simulated dwarfs in groups and clusters in TNG50 (gray dots) and a compilation of observational data (purple symbols) including normal dwarfs \citep[translucent purple shapes][]{Forbes2018, Peng2008, Prole2019, Lim2018} and UDGs \citep[filled purple shapes][]{Gannon2022, amorisco2018, vanDokkum2017, Saifollahi2021, Lim2018, Lim2020, Somalwar2020}. We highlight simulated UDGs in TNG50 with orange empty circles, which we compare to observed UDGs shown in solid purple.  

Fig.~\ref{fig:abundance} indicates that simulated UDGs display GC numbers that overlap well with the majority of available observations of UDGs (left panel), including systems in low mass groups \citep{Somalwar2020} but also high-density environments like Coma \citep{Amorisco2016, Gannon2022}. We note, however, that extreme UDGs with $N_{\rm GC} > 30$ are not present in our simulated catalog but seem to be present in observations. 

This result is not entirely unexpected: all UDGs in TNG50 populate dwarf halos in the mass range $M_{vir} = [2\times10^9, 2\times10^{11}]$ \msun at infall \citep[using the last time a halo is a central as definition of infall time,][]{Doppel2023}, and their GC content is a reflection of this prediction. The specific frequency of GCs for these galaxies is shown on the right panel of  Fig.~\ref{fig:abundance} and confirms a similar trend: while there is good overlap for many of the simulated UDGs in TNG50, very extreme values with $S_N \gtrsim 50$ are not produced in our simulated sample but exist in systems like the Virgo or Coma cluster \citep{Lim2018, Lim2020}.

\begin{figure}
    \centering
    \includegraphics[width = \columnwidth]{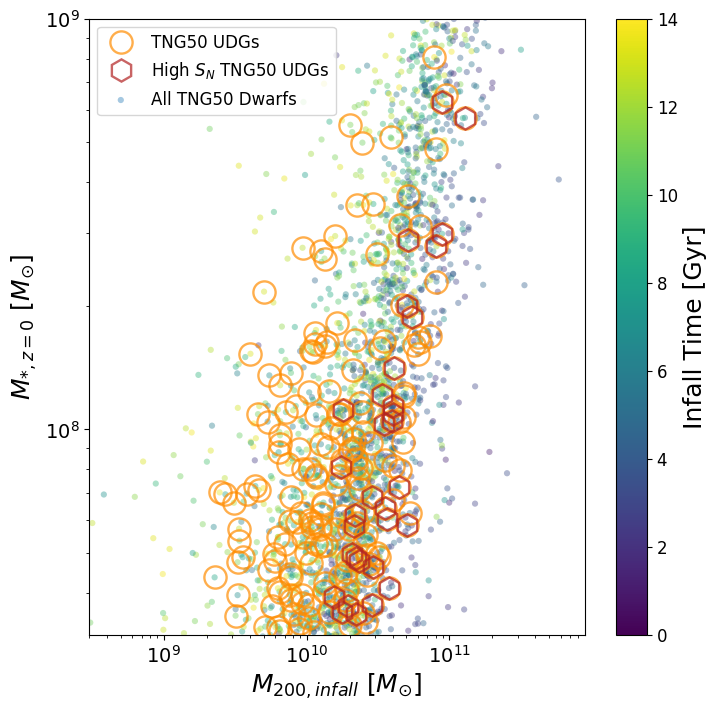}
    \caption{Stellar mass at $z = 0$ ($M_{*, z = 0}$) vs. virial mass at infall ($M_{200, infall}$) for satellite galaxies in the mass range explored. Symbols are color-coded by their infall time in Gyr, such that yellow colored points correspond to a recent infall and bluer points correspond to an earlier infall.
    UDGs are highlighted by orange circles. We highlight with red hexagons UDGs with the highest $S_N$ in our sample (top 15\% of $S_N$ at fixed $M_V$). These more extreme UDGs tend to have earlier infall times and more massive halos than their less extreme counterparts.}
    \label{fig:infallmvir}
\end{figure}

Identifying GCs that are associated to a given galaxy in observations is not without challenge, a subject we return to in Sec.\ref{sec:mock}. The iconic UDG DF44 is a good example \citep{vanDokkum2016}. Originally thought to host nearly $100$ GCs \citep{vanDokkum2016}, it has been now estimated to have only $\sim 20$ GCs \citep{Saifollahi2021}. If we take the latest measurements as correct, our simulated UDGs are a good representation of galaxies like DF44. On the other hand, if earlier estimates are found to hold, then we do not find DF44 analogs in our sample. The example set by DF44 perhaps warrants a closer look into observed galaxies with very extreme GC content.

Despite the lack of direct analogs to the most extreme observed UDGs in terms of GC number, simulated GC systems encouragingly span a relatively wide range of GC contents, in good agreement with observational claims \citep[e.g., ][]{Lim2018, Lim2020, Toloba2023}. Of particular interest are those with the largest numbers of GCs (or specific frequency) at any given mass (or luminosity). A closer look to the set of TNG50 UDGs in the top $15\%$ of GC number and specific frequency at fixed stellar mass (and $M_V$) reveal that these UDGs tend to reside in higher mass---albeit still dwarf-mass---halos at infall (Fig.~\ref{fig:infallmvir}, where high $S_N$ UDGs are highlighted in red).

Interestingly, this bias towards higher mass halos for more extreme UDGs is linked to earlier infall times than their less extreme counterparts. This is illustrated clearly with the color coding of symbols in Fig.~\ref{fig:infallmvir}. This finding is similar to our previous results exploring the GC content of normal dwarfs in the Illustris simulations \citep{ramosalmendares2020}. More specifically, at fixed $z=0$ stellar mass, galaxies with early infall times are biased towards higher halo mass due to the time evolution of the $M_*-M_{halo}$ relation with redshift. Larger halo masses imply a larger number of GCs assigned at infall. In addition, galaxies that infall early stop forming stars longer ago, meaning that they have passively evolved their stellar population becoming fainter in V-band magnitude and consequently increasing their specific frequency. In TNG50, we find a median infall time $t_{\rm inf} \sim 6.1$ Gyr for our large GC content UDGs compared to $t_{\rm inf} \sim 8.1$ Gyr for the rest of the UDG sample. 

\begin{figure}
    \centering
    \includegraphics[width = \columnwidth]{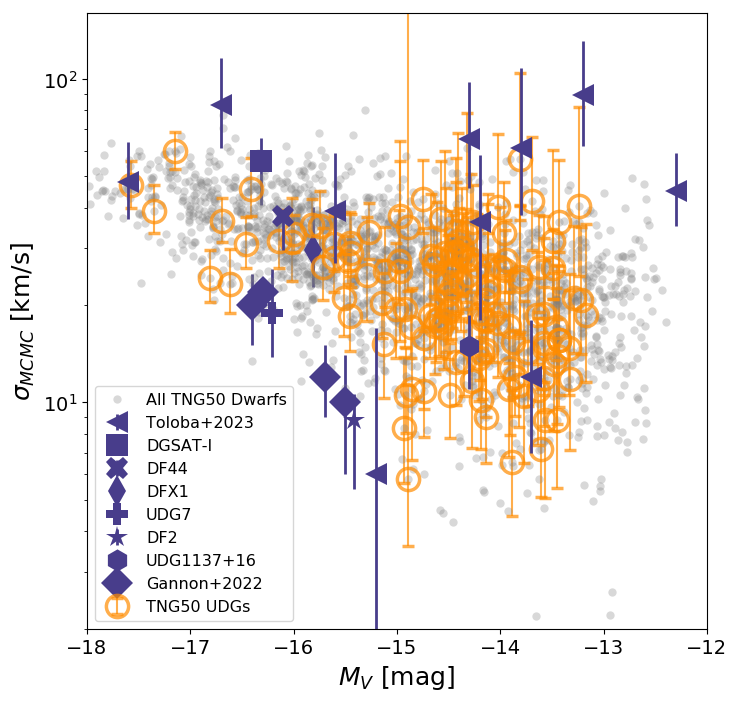}
    \caption{Kinematics of the GC systems of dwarf galaxies in TNG50 calculated via a Markov-Chain Monte Carlo (MCMC) method with as Jeffrey's prior plotted against host galaxy V-band magnitude, $M_V$. UDGs in TNG50 are highlighted with orange circles with errorbars representing the 25-75 percentiles from the PDF generated stochastically by the MCMC method. All dwarf satellites from TNG50 are shown as gray points. We show observations of GC kinematics from UDGs coming from various studies as large, solid, purple shapes. We find a wide range of UDGs represented in the literature, with some having dispersions that put them in the range of ``normal'' dwarf galaxies, some with dispersions that put them in the dark matter deficient category. Other observed UDGs sit above what is predicted by TNG50, suggesting that they reside in rather overmassive halos. We note that much of the scatter $\sigma_{MCMC}$ for the UDGs in TNG50 is due to the presence of few GC tracers, making many of the lower scattering points the product of small number statistics. UDGs and their GC systems in TNG50 thus appear to be kinematically indistinguishable from normal dwarf galaxies. Large $\sigma$ values seem underrepresented in our sample compared to measurements in the Virgo cluster \citep{Toloba2023}.}
    \label{fig:kinematics}
\end{figure}

\jed{As with} GC content, the velocity dispersion of observed UDGs has been shown to span a wide range. From the popular DF2 and DF4  galaxies associated to NGC1052, whose velocity dispersions ($\sigma < 10$ km/s) are so low that they are consistent with no dark matter at all \citep[e.g., ][]{vanDokkum2018, Danieli2019} to UDGs nearing $\sigma \sim 100$ km/s, compatible with halos so massive that could in theory host MW-like galaxies. Of particular interest is the recent study by \citet{Toloba2023}, which represents the first \textit{systematic} study of the GC kinematics of UDGs in the Virgo cluster. Half of their sample ($5$ out of $10$) shows velocity dispersion $\sigma \geq 50$ km/s measured within $1.5$-$2$ kpc projected radii, making them consistent with inferred halo masses $M_{halo} \geq 10^{12}$\msun---on par with that of the MW (see Fig. 9 from \citet{Toloba2023}). The authors also report at least one UDG that is also consistent with having no dark matter, which seems to be tied to the ongoing tidal disruption of that particular UDG, partially explaining some of the diverse $\sigma$ values in the sample.

We show the measurements presented in \citet{Toloba2023}, along with a compilation of other available velocity dispersions for observed UDGs in Fig.~\ref{fig:kinematics} (purple shapes). The GC velocity dispersion of simulated UDGs in TNG50 are shown with unfilled orange circles. Following \citet{doppel2021}, we have estimated GC velocity dispersion for these systems following an Markov-Chain Monte Carlo (MCMC) method with a Jeffreys prior on the dispersion itself, as this method was found to be the most adequate to estimate $\sigma$ with a small number of tracers. The error bars on the orange circles show the $25\%$-$75\%$ spread in the velocity dispersion from the PDF stochastically generated via the MCMC method. This is analogous to the way that velocity dispersions were calculated for the GC systems of Virgo-cluster 
UDGs \citep[][among others]{Toloba2023}. We include the dispersion of other UDGs in the literature derived from GC kinematics \citep[NGC1052-DF2, ][]{vanDokkum2018}, stellar kinematics \citep[DF44, ][]{vanDokkum2019}, and stellar spectra (DFX1 \citep{vanDokkum2017}, DGSAT-1 \citep{MartinexDelgado2016, MartinNavarro2019}, UDG7 \citep{Chilingarian2019}, UDG1137+16 \citep{Gannon2021}, and UDGs from the Perseus cluster \citep{Gannon2022}).
This set of observed UDGs are selected here to be all consistent with the UDG definition presented by \citet{Lim2020}, in that they are outliers of more than $2.5\sigma$ in one of the scaling relations between luminosity and surface brightness, mean effective surface brightness, and effective radius.

Encouragingly, the range of GC velocity dispersions predicted by the tagged GCs in TNG50 agrees well with the bulk of observed values for UDGs, in particular for objects with normal-dwarf velocity dispersions such as DFX1, UDG7, UDG1137+16, several Virgo UDGs, and DF44. About half of the UDGs with available velocity measurements are consistent with a dark matter content of a dwarf-mass halo---in agreement with predictions from our UDG sample in TNG50. Moreover, the GC velocity dispersion of simulated UDGs overlaps well also with non-UDG dwarf satellites in TNG50 (gray dots). This is indeed expected from the formation scenario of UDGs in this simulation, which place them in dwarf dark matter halos consistent with the non-UDG sample \citep[although with a small bias towards higher mass, e.g.,][]{Benavides2023}.

Interestingly, we also see in Fig.\ref{fig:kinematics} several UDGs and dwarfs from TNG50 that show $\sigma_{\rm MCMC}<10$ km/s, reminiscent of dark-matter free UDGs such as {\color{violet}NGC1052-}DF2. A closer inspection of this simulated analogs to {\color{violet}NGC1052-}DF2 show that several have undergone a rather significant amount of dark matter stripping \citep[as was found in ][]{doppel2021}. However, much of the scatter in the lower $\sigma$ UDGs arises from having only 3-5 GCs to recover the potential of their host halo. As \citet{doppel2021} showed, using a Jeffrey's prior for a low number of tracers performed well in recovering dynamical mass in the \textit{median} of the sample, but with a large galaxy-to-galaxy scatter. This is a large contributor to the source of kinematic analogs to NGC1052-DF2 in TNG50 and highlights the importance of having a sufficient number of tracers to make accurate {\it individual} dark matter mass estimates.

\begin{figure*}
    \centering
    \includegraphics[width = \textwidth]{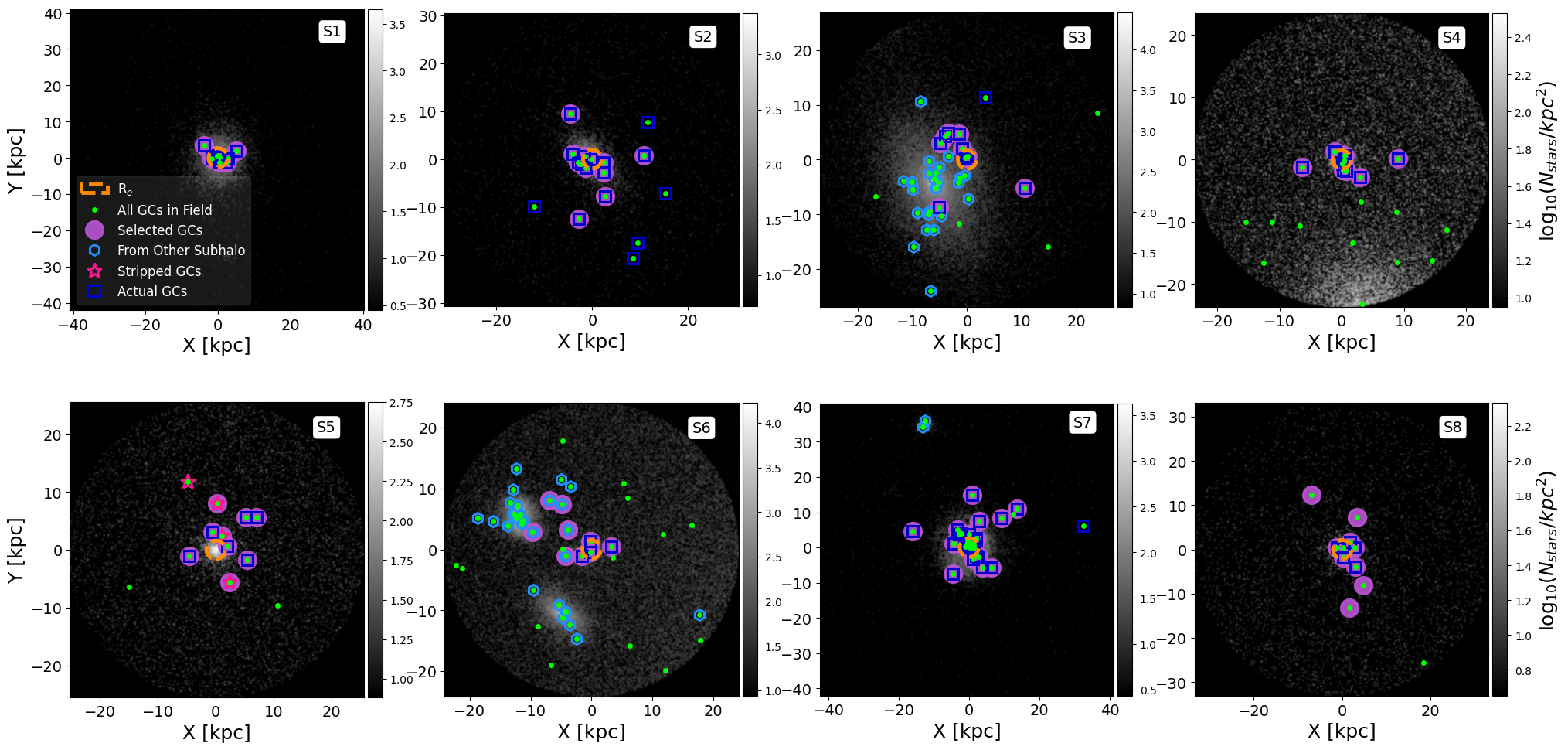}
    \caption{Mock X-Y projections of stars (background grayscale, colored by the number density of stars in each bin) and GCs (lime green points) within $16 R_e$ of 8 UDGs within TNG50. We name the satellites S1-S8 as annotated in the upper right corner of each panel. The UDGs shown are selected to have at least 8 GCs in within $16 R_e$ of the host UDG and to display a range of scenarios from quite easy to surprisingly difficult for selecting bound GCs (see Sec.~\ref{sec:mock}). GCs that would be considered associated in observations are highlighted with an underlying large purple circles, those that belong to other subhalos by sky blue hexagons, those that are tidally stripped by pink stars, and actual GCs bound to the subhalo by dark blue squares. For reference, we show the $R_e$ of each UDG as dashed, orange circles. Several UDGs, namely S1, S2, S5, and  S8 are quite isolated, with the rest having one or more other galaxy in the field of view. From spatial information alone, determining GC boundness is not straightforward.} 
    \label{fig:projections}
\end{figure*}

On the other hand, UDGs with high GC velocity dispersion, $\sigma_{\rm MCMC} > 50$ km/s, are less common in our simulated sample compared to available observational constraints. A closer inspection of our high-velocity cases shows a similar situation as described above: they tend to have 3 - 5 dynamical tracers and scatter upwards of their true velocity dispersion (as measured from their mass content within $R_e$). High-dispersion objects are interesting because they do not conform to the expectations of dark matter content given their luminosity. Several candidates have been hinted at in observations including, for example, objects like DGSAT-1 \citep{MartinexDelgado2016} and NGVSUDG-09, NGVSUDG-05, NGCSUDG-11, NGVSUDG-19, NGVSUDG-20 and NGVSUDG-A04 from the \citet{Toloba2023} study of UDGs in Virgo. These are often interpreted as ``failed'' massive halos that were destined to form a galaxy more comparable to the Milky Way, but stopped forming stars much earlier than expected, resulting in an overly-massive halo given its stellar mass \citep{vanDokkum2015a,Peng2016, vanDokkum2017, Lim2018, Lahen2020, Danieli2022, Janssens2022}. Calculations presented in \citet{Toloba2023} show that halos more massive than $M_{200} \sim 10^{12}$\msun\; are necessary to explain the kinematics of the large-$\sigma_{\rm MCMC}$ UDGs.  Such ``failed'' galaxies are not present in the simulated UDG sample in TNG50.

This finding may have different explanations. The most straight-forward one is that there may be a legitimate disagreement between theory and observation, implying that the physical mechanisms to form such massive failed galaxies is missing from cosmological simulations (as no other simulation has reported successfully forming such dark matter dominated objects to date) and from our understanding of galaxy formation. Alternatively, the origin of the large velocity dispersion in observed UDGs may be attributed to the presence of observational errors (which are not considered in Fig \ref{fig:kinematics}), interlopers and/or observational biases which are not currently included when comparing with theoretical predictions. We use our simulated GC catalog to more closely address whether contamination alone may explain the observed UDGs with large inferred dark matter halo masses.

\section{Effects of Interlopers on the GC velocity dispersion of UDGs}
\label{sec:mock}

\begin{figure*}
    \centering
    \includegraphics[width = \textwidth]{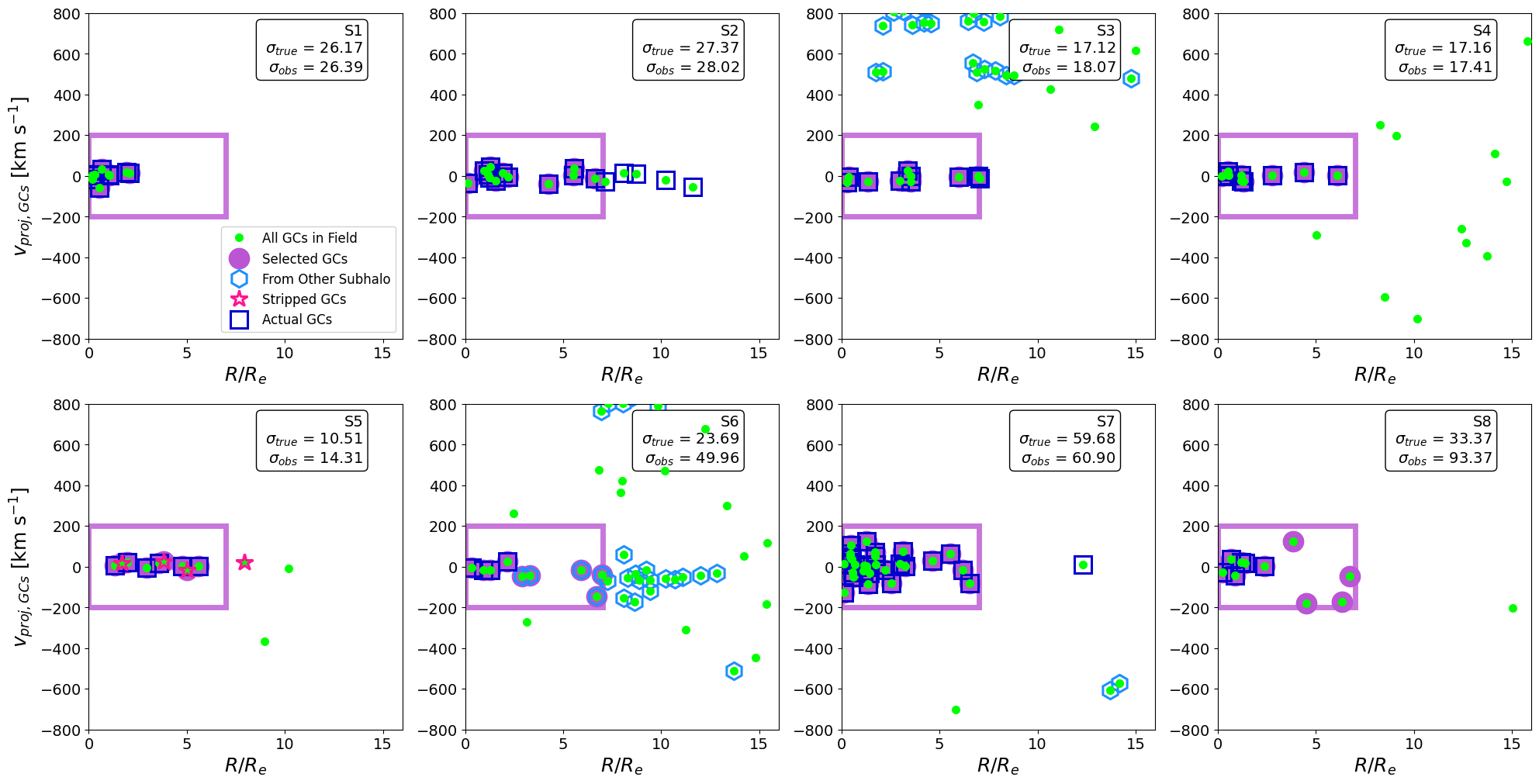}
    \caption{A mock observation of the radial velocity of the GCs associated to the eight UDGs in Fig. \ref{fig:projections}. GCs are considered members of the galaxy if they fall within $7R_e$ of a given galaxy and their radial velocities are within $\pm 200 km/s$ of that of their suspected host galaxy. The wide range in acceptable velocities allows for the possibility of detecting UDGs that form in massive dark matter halos. All GCs in the $16 R_e$ field of view are represented by lime green dots, with those selected to be associated by the radius and velocity cuts are highlighted by large purple circles within the purple box. GCs that have been tidally stripped from their host UDG are outlined by magenta stars, with GCs known to belong to the UDGs highlighted by unfilled dark blue squares and those that belong to other subhalos in the field of view are outlined by sky blue hexagons. We can see that even in the presence of additional galaxies in the field of view, such as S3 that the actual GCs can be recovered by the selection method described in Section ~\ref{sec:mock}. For some UDGs (namely S3, S4, and S7 in the set of UDGs presented here), we find the observed radial cut of $7 R_e$ to be somewhat conservative. Overall, assigning GCs based on kinematics is overall a powerful tool, with a nearly correct set of GCs often being picked out of crowded fields (e.g., S3). Interestingly, GCs that are considered part of the ICGCs that have been tidally stripped from their hosts are often difficult to distinguish kinematically or radially from the set of actual GCs, although it does not seem to affect the estimate of velocity dispersion (see S5). Overlap of several galaxies in the field of view can complicate GC identification (see S6), but cases like this would likely not be included in observational samples. S8 represents an interesting case in which interloper GCs from the intra-cluster component are flagged as members and substantially increase the estimated velocity dispersion.}
    \label{fig:vproj}
\end{figure*}

The analysis of the simulated UDGs and their GCs in Sec.~\ref{sec:gc_abundance} assumes that only the gravitationally associated GCs are taken into account when estimating GC numbers and kinematics. For the case of the TNG50 simulations, we use information from Subfind to determine whether or not a GC is gravitationally bound to a given UDG. However, this is not possible in observations,
where assigning membership to GCs nearby a galaxy of interest becomes an additional challenge.

In the specific sample from the Virgo cluster, where most of the available kinematical constraints on UDGs exist \citep{Toloba2018, Toloba2023}, GC membership is based on a combined criteria in projected distance to the host galaxy: $R < 7 R_e$, with $R_e$ the effective radius of the host UDG, and an additional restriction on the relative line-of-sight velocity between the candidate GC and the UDG, set to be less than $200$ km/s. We can use our simulated catalogs to evaluate the degree to which the selection effects and specific choices applied in observed samples may lead to the possible inclusion of interloper GCs, biasing the velocity or mass estimate for some UDGs. 

We construct mock observations of our simulated samples by projecting all groups and clusters in a given direction and applying a similar selection criteria as described in \citet{Toloba2023}. By doing so, we are considering the top two possible contamination sources: $i)$ GCs associated to other galaxies that are near the UDG in projection and $ii)$ GCs in the diffuse intra-cluster GC component (ICGCs). 
Assuming that the luminous mass of the UDGs is distributed roughly spherically, we make the conversion between 3D stellar half-mass radius ($r_{h,*}$ and projected effective radius ($R_e$) using $R_e = 3/4 r_{h,*}$ \citep[e.g.][]{hernquist1990, Wolf2010, Somerville2018}.

For illustration, Fig.~\ref{fig:projections} shows $8$ representative examples of simulated UDGs and their GCs in our sample. The stellar number density of the UDGs and their surroundings is shown by the background grayscale, and the GCs that fall in projection within the frames are represented by different symbols (see legend). We label them satellite-1 through -8, or S1-S8 for short, with a label on the upper right-hand corner of each panel. We can find UDGs in relatively isolated surroundings (such as S1, S2, S5, and S8) as well as to those in crowded or obviously with interlopers from several companion galaxies in projection (S3, S4, S6, and S7). These examples are chosen to showcase different levels of contamination by interlopers and are not a random selection of UDGs in our sample.

\begin{figure}
    \centering
    \includegraphics[width = \columnwidth]{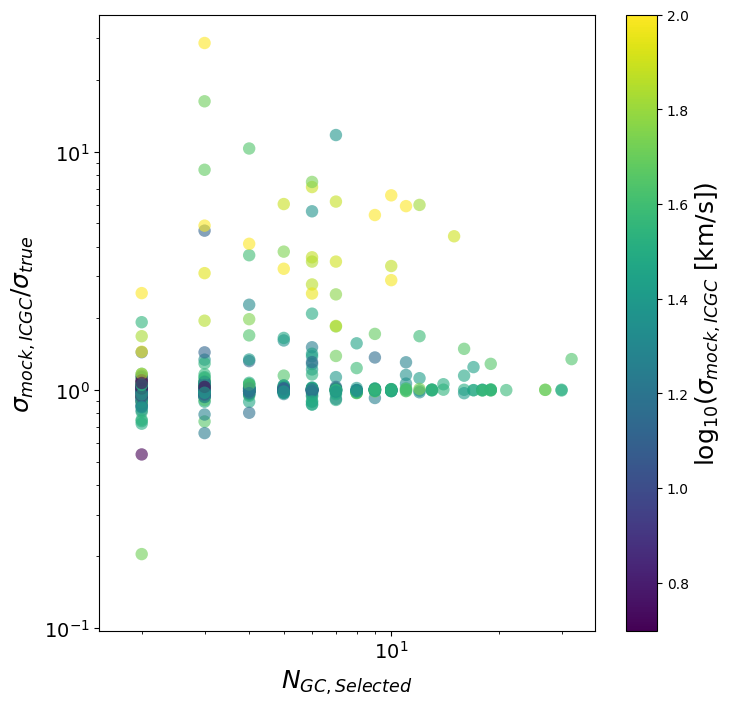}
    \caption{The ratio of the GC velocity dispersion measured via mock observations, $\sigma_{mock, ICGC}$. to the actual GC velocity dispersion, $\sigma_{true}$ as a function of total GCs selected, $N_{GC, Selected}$ via the method described in sec. \ref{sec:mock}, with points colored by $\log_{10}(\sigma_{mock, ICGC})$. 
    We can see that for small GC numbers, the $\sigma_{mock, ICGC}$ can be greatly inflated from its true value, especially from intra-cluster GC contaminants. For most galaxies, the mock observations do not pick up a significant number of interloping intra-cluster GCs in the mock observations, leading to an overall median $\sigma_{mock, ICGC}/\sigma_{true}\sim 1$.} 
    \label{fig:icgc_contam}
\end{figure}

Next, we apply the selection criteria in GC radial velocity, $v_{proj, GC}$. Fig.~\ref{fig:vproj} shows this for the $8$ examples discussed above. For convenience, we center the GC velocities on that of the host UDG. Following \citet{Toloba2023}, we consider GCs within $7R_e$ of their host galaxy and within $\pm 200$ km/s of the velocity of their host galaxy as bound to the host galaxy (purple box). GCs that would be selected as members by this method are lime green dots highlighted by large purple circles, while those outside of the selection box are shown in lime green. 

We use our simulation to obtain additional information for each GC. Those known to be gravitationally bound to the UDGs (based on SubFind information) are outlined by dark blue squares. GCs that belonged to the UDG but have now been tidally stripped are outlined by magenta stars, and those outlined by sky blue hexagons are GCs associated to other subhalos. Lime green dots without any outlining shape belong to the intra-cluster GC component. In all panels we quote, on the upper right corner, the actual 1D velocity dispersion calculated with all bound GCs ($\sigma_{\rm true}$) along with the corresponding velocity dispersion computed using the objects within the selection box ($\sigma_{\rm obs}$). We emphasize that, similar to observational samples, the velocity dispersion determination is computed using an MCMC method assuming a Jeffreys prior. 

In general, we find that this simple selection criteria works rather well in most cases considered, with a few exceptions. We can see that for all eight featured UDGs, most of the GCs gravitationally bound to the galaxy are recovered by this selection method, with the exception of S2 and S7, which are missing $5$ and $1$ associated GCs, respectively, when the selection criteria are applied. Note that in neither case does this matter for the velocity dispersion measured, which remains very close to the true value even when missing a few GCs (upper right corner of each panel). 

As expected, the inclusion of velocity information is critical to remove GC interlopers. For example, S3 and S6 in Fig.~\ref{fig:projections} have obvious contamination ongoing due to the overlap in projection with other satellites in the group. We can see in Fig.~\ref{fig:vproj} that the addition of velocity removes the interlopers associated with S3. However, this is not the case for S6, where GCs bound to the companion galaxy fulfill the criteria of membership due to chance alignment in the velocities. This results, for the specific case of S6, in a factor $2$ overestimation of the velocity dispersion inferred: using the GCs within the selection box results in $\sigma_{\rm obs} \sim 50$ km/s whereas the truly associated GCs are moving with $\sigma_{\rm act} \sim 24$ km/s.

While the case of S6 demonstrates that care must be exercised when dealing with projected data, it presents a type of contamination that observational studies will avoid unless absolutely necessary. In fact, none of the UDGs considered in the sample of \citet{Toloba2018} or \citet{Toloba2023} contains other galaxies in projection on the line of sight that are brighter than $M_V \sim -13$; therefore, they are not luminous enough to have GCs that pose the risk of significantly contaminating the GC sample \citep[see sec 5.1 of ][]{Toloba2023}. In what follows, we choose to ignore contamination from GCs associated to other subhalos, as observational studies would purposely remove such complicated systems from their samples. 

However, a more subtle case is that of S8 in our sample. S8 is seemingly isolated, but several intra-cluster GCs fall within the selection box, artificially enhancing the velocity dispersion measured by a factor of $\sim 3$. This galaxy would be inferred to inhabit a massive dark matter halo with $\sigma_{GC} \sim 100$ km/s, while in reality 
it inhabits a dwarf-mass one with $\sigma_{\rm true} \sim 35$ km/s. This presents a concrete example where an otherwise relatively normal UDG could be kinematically mistaken as bearing an overly-massive halo. 

Are cases like S8 common in our sample? For that, we need to evaluate how often contamination from the intra-cluster component sneaks into the selection box. We quantify this in Fig.~\ref{fig:icgc_contam}. We show, as a function of the number of GCs within the selection box in our UDGs, $N_{\rm GC, \rm Selected}$, the ratio of the measured velocity dispersion (including intra-cluster interlopers) and the true value (computed with only bound GCs according to SubFind). For the vast majority of simulated UDGs the velocity dispersion estimate remains within $20\%$ of its true value, suggesting that it is not likely that interlopers will play a dominant effect in the majority of UDG measurements. However, for systems with less than $10$ GCs, the inclusion of intra-cluster contamination may cause overestimation of the velocity by factors $2$-$10$. The median and percentiles show, however, that it is statistically much more likely to remain within $15\%$ of the true value.

\begin{figure}
    \centering
    \includegraphics[width = \columnwidth]{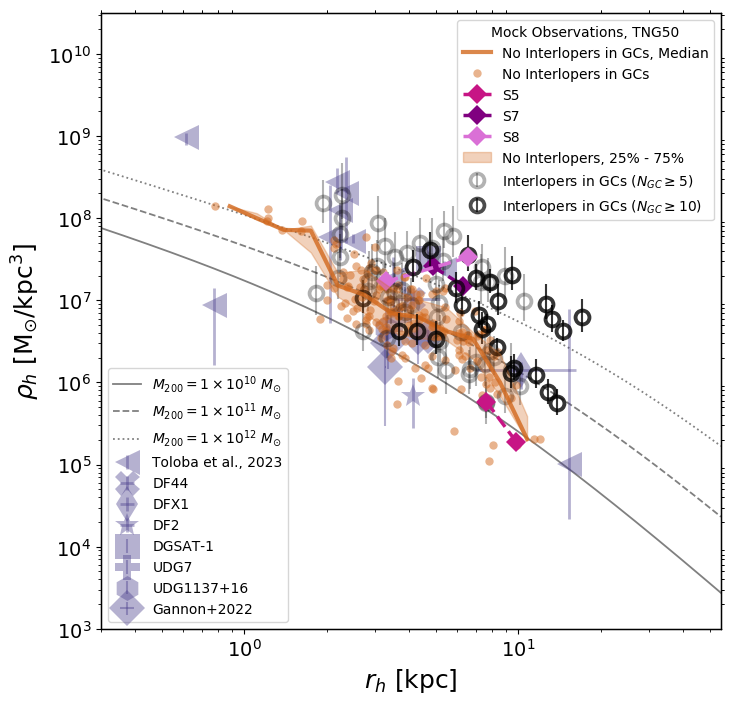}
    \caption[Effects of GC contamination on dark matter mass inferences]{Effects of GC contamination on dark matter mass inferences. We show the mass density, $\rho_h$, as a function of radius as reconstructed from the kinematics of GCs for UDGs with more than 5 GCs in projected mock observations (see Sec. \ref{sec:mock} for details). To guide the eye, gray lines show the density profiles of NFW halos with concentration $c = 10$ and halo masses $M_{200} = 10^{10}, 10^{11}, 10^{12}$\msun\; are shown as the solid, dashed, and dotted gray lines respectively. The instantaneous density at the half-number radius of GCs is represented by brown dots for UDGs with no GC interlopers. The thick brown line and shaded region indicates the median and 25-75 percentiles of this distribution and recovers well the $M_{200} = 10^{11}$\msun\; density profile. GCs are good tracers of mass for these systems. However, contamination by intra-cluster GCs seemingly associated in projection cause a larger scatter, including several systems laying close to the $M_{200} = 10^{12}$\msun\; line (gray symbols for more than $5$ GCs associated in projection, black for more than 10, each system considered in XY, XZ and YZ projection). Error bars in simulated UDGs correspond to the 25\%-75\% scatter from their MCMC pdfs. Values compiled from the literature for real UDGs are shown in purple symbols as labeled. 
    Specific cases of contamination from Fig.~\ref{fig:vproj} and \ref{fig:projections} are highlighted in colored diamonds, connecting their density calculated using their actual GCs and those selected via observational methods. The presence of intra-cluster GC interlopers helps explain the inference of overly-massive halos in some of our simulated UDGs, for example in our example S8.}
    \label{fig:mockobs}
\end{figure}

\subsection{Can intra-cluster GCs then explain the high-incidence of large velocity dispersion UDGs found in Virgo?} 

A close inspection of Fig.~\ref{fig:vproj} shows that interlopers tend to have the largest distances and largest velocity difference with the central galaxy (yet still remaining within the selection box). We have re-analyzed the velocity dispersion of the most extreme ``failed galaxies" example in Virgo from \citet{Toloba2023} removing the furthest GC or the largest velocity difference GC and found no significant change in the estimates of their velocity dispersion or dynamical mass. These include NGVSUDG-05, NGVSUDG-09, NGVSUDG-11, NGVSUDG-19, NGVSUDG-20, and NGVSUDG-A04 using the nomenclature of the original paper. The most extreme variation is for NGVSUDG-09, which changes from $\sigma = 83^{33}_{-22}$~km/s to $\sigma = 60^{25}_{-15}$~km/s. While these values are still statistically consistent, the median velocity dispersion is brought more in line with TNG50 UDGs.
Worth noticing, NGVSUDG-19 has only $3$ GCs members identified, so it is necessary to proceed with caution regarding this particular target.

In order to evaluate the possibility of contamination in the \citet{Toloba2023} sample more closely, we restrict now our simulated sample to only UDGs outside of $0.1R_{vir}$ from their host cluster and with $N_{\rm GC,Selected} \geq 5$, (only excluding 1 target from \citealt{Toloba2023}). A total of 242
UDGs satisfy these criteria when using $3$ different projections---along the $x$-, $y$- and $z$- axis of our $39$ groups and clusters in TNG50. We derive from these mock projections: the corresponding 1D MCMC velocity dispersion, the half-number radius of the GCs, and the dynamical mass at half-number radius following Jeans modeling as in \citet{Wolf2010}. The dynamical mass may be trivially transformed into mass density by dividing by the spherical volume enclosed within the half-number radii.

Fig.~\ref{fig:mockobs} shows the inferred mass density for this subsample of simulated UDGs as a function of the half-number radius of their selected GCs. Simulated galaxies with $5$-$9$ selected GCs with at least one interloper are shown in gray, unfilled, circles while those with 10 or more (including interlopers) are indicated by black, unfilled, circles. For reference, the gray lines represent NFW profiles with virial masses $M_{200} = 10^{10}, 10^{11}$ and $10^{12}$\msun\; and a concentration $c=10$. We do find that for most cases, in particular those with 10 or more GCs, galaxies cluster around the $10^{11}$\msun\; line, which is in agreement with the prediction in TNG50 that UDGs occupy dwarf-mass halos with $M_{200} \sim 10^9 - 10^{11}$ \msun \citep{Benavides2023}. 
For comparison in Fig. \ref{fig:mockobs}, we include the inferred density of several observed UDGs that are derived from their reported velocity dispersions (whether from GCs, stars, or stellar spectra) and half number radii (GCs) or effective radii (stars) via dynamical mass estimation \citep[see][]{Wolf2010}.

However, we find several instances where the mocked galaxies wander close to or above the $M_{200} \sim 10^{12}$\msun\; line, despite their true halo mass being substantially smaller. This happens mostly due to contamination from interloping intra-cluster GCs. This can be appreciated relative to the UDGs that possess no interloping GCs in their mocked GC sample, as shown by the brown points, line (median), and shaded region ($25\%-75\%$ spread) in Fig \ref{fig:mockobs}. There is a much larger incidence of gray circles near the $M_{200} \sim 10^{12}$\msun\; line, suggesting that low numbers of kinematical tracers may play a role in the appearance of overmassive halos.
Such is the case of S8 introduced before in Figs.~\ref{fig:projections} and \ref{fig:vproj}, highlighted in pink, which moves from a true mass-density consistent with the dashed line ($M_{200} \sim 10^{11}$\msun) to its inferred density more consistent with a MW-mass halo with $M_{200} > 10^{12}$\msun.

Worth discussing is also the case of S7, highlighted in Fig.~\ref{fig:mockobs} as the purple diamond. As shown in Fig.~\ref{fig:vproj}, S7 does not include contamination by GC interlopers in its mocked GC sample. Yet its inner density is high and consistent with MW-like halos both when applying the mock selection in projection or when considering all bound GCs according to Subfind. We have checked that this high density is not the result of an overly-massive halo but instead corresponds to a dwarf-mass halo with a larger-than-typical concentration. The virial mass before infall for S7 is $M_{200} \sim 9 \times 10^{10}$\msun. This galaxy is a good reminder that variations in concentration may also drive some of the scatter in the inferred dark matter content of UDGs, a possibility briefly discussed in \citet{Gannon2022}.

Given these results, can intra-cluster GCs explain the high-incidence of large velocity dispersion UDGs found in Virgo? Within the range of galaxy groups and clusters that we can explore with TNG50, we find that contamination from intra-cluster GCs is unlikely to explain the high incidence of high-mass UDGs in Virgo reported recently by \citet{Toloba2023}. Only a handful of simulated UDGs are driven close to the $M_{200} \sim 10^{12}$\msun\; line due to contamination effects, with only $9.5\%$ of UDGs with 5 GCs or more showing velocity dispersion overestimation by a factor of 2 or more in the mock observations. However, a factor to keep in  mind is that even the most massive simulated galaxy cluster in TNG50 (Group 0, $M_{200} = 1.87 \times 10^{14}$\msun) is on the the low end of mass estimates for the Virgo cluster \cite[$M_{200} \sim 2-9 \times 10^{14}$ \msun,][]{Weinmann2011, karachentsev2010}, with the remainder of our groups in the simulated sample being lower mass. For our most massive cluster, we predict a total of 34,231 GCs with $M \geq 7 \times 10^3 \; \rm M_\odot$, which is on par with what is expected for the GC number density of the M87 subgroup in Virgo \citep[e.g., ][]{durell2014, lee2010}, but is about a factor of two lower than the \textit{combined} estimate when considering also the M49 subgroup, $N_{GC, Virgo} \sim 67,300 \pm 14,400$ \citep{durell2014}. All of the remaining groups in our simulated sample are less massive and will therefore have less GCs than Group 0. It is therefore possible that chance alignment of ICGCs has a larger impact in the specific case of observations in Virgo than found on average in our study.

\section{Discussion and Summary}
\label{sec:summary}

We use a catalog of GCs added to the TNG50 cosmological simulation, introduced in \citet{Doppel2023}, to study the population of GCs associated to UDGs with stellar mass $M_*=[10^{7.5},10^9]$\msun\;
in $39$ groups and clusters with $M_{200}=[5 \times 10^{12} \rm - 2 \times 10^{14}]$\msun. UDGs are selected as outliers in the mass-size relation as presented in \citet{Benavides2023}. 

UDGs in TNG50 are found to form in dwarf-mass halos with biased-high spins and virial masses between $M_{200} = [2\times10^{9}, 2\times10^{11}]$\msun. As a result, simulated UDGs have similar GC numbers to those associated with non-UDG dwarfs of similar stellar mass. We find between 1-30 GCs bound to the simulated UDGs, with only 12 UDGs having no GCs at all. This seems in agreement with observed UDGs, which show a large spread in GC content \citep{amorisco2018, Lim2018, Lim2020, Somalwar2020, Gannon2022, LaMarca2022, Toloba2023}. 

However, our sample lacks extreme outliers, with $N_{GC} > 30$, and $S_N > 50$, as some observations suggest \citep[e.g.][]{Peng2016, Lim2018, Lim2020, Muller2021}. The lack of high specific GC frequency simulated UDGs is ultimately linked to the fact that UDGs in TNG50 all inhabit dwarf-mass halos, which have low GC numbers according to the scaling assumed in the model.  We caution, however, that uncertainties are still important in observations. For example, our predictions fall well below the initial number of $\sim 100$ GCs reported for the iconic DF44 \citep{vanDokkum2017} but agree very well with its revised value $\sim 20$ in the more recent work by \citet{Saifollahi2021}.

As for the GC numbers, we find in general good agreement between the predicted GC velocity dispersion in simulated UDGs and values reported in the literature for observational samples. Our predictions agree well with $\sigma$ measurements for a number of UDGs, particularly DF44, DGSAT-1, DFX1, UDG7, and several UDGs in the Virgo cluster. However, large velocity-dispersion outliers with $\sigma > 50$ km/s such as those found for half of the UDGs studied in the Virgo cluster in \citet{Toloba2023} are not common in our sample. 

We can use our simulated GC catalogs to make projected mock observations of our systems and assess whether interloper GCs could affect the observational results. We find that outliers from the intra-cluster GC component associated to the host galaxy group or galaxy cluster may in some cases impact the velocity dispersion measurement, inflating $\sigma$ by factors of $>2$. These cases are, however, rare, in particular when focusing on UDGs with a sufficient number of tracers (GCs).

In agreement with our previous results \citet{doppel2021}, we find $10$ or more GCs are needed for robust kinematical measurements. For instance, only $9.5\%$ of cases with more than $10$ GCs have velocity dispersions that are overestimated by more than a factor of $2$ because of the presence of interlopers. Such cases will suggest dark matter halos with $M_{200} \sim 10^{12}$\msun when in reality they occupy normal dwarf-mass halos. 

We compare our results with the high incidence of observed UDGs with large velocity dispersions reported in kinematical studies of UDGs in the Virgo cluster and conclude that the frequency of contamination in our systems does not explain the large number of UDGs with $\sigma > 50$ km/s in Virgo. A caveat of our study is that groups and clusters included in TNG50 are on average less 
 massive than Virgo, and the incidence of interloper contamination could be higher in more massive systems. We identify some high inferred halo-mass cases in \citet{Toloba2023}, such as UDG 19 or 05 and 20, that have 5 GC tracers or less, making them interesting candidates to follow
up spectroscopically for confirmation. Ultimately, for UDGs with a low number of identified GC members, measuring their stellar velocity dispersion might be the only avenue to constrain better their true dark matter mass content and, with it, their possible formation path.

\section*{Acknowledgements}

JED and LVS are grateful for financial support from the NSF-CAREER-1945310 and NASA ATP-80NSSC20K0566 grants. ET is thankful for the support from NSF-AST-2206498 and HST GO-15417 grants. DN acknowledges funding from the Deutsche Forschungsgemeinschaft (DFG) through an Emmy Noether Research Group (grant number NE 2441/1-1).

\section*{Data Availability}

The realistic GC catalogs used in this study are available to the public. The catalogs can be downloaded here: www.tng-project.org/doppel22 or as part of the TNG50 public data release \citep{nelson2019dr}.



\bibliographystyle{mnras}
\bibliography{gc_paper_udgs} 

\begin{thebibliography}{}
\makeatletter
\relax
\def\mn@urlcharsother{\let\do\@makeother \do\$\do\&\do\#\do\^\do\_\do\%\do\~}
\def\mn@doi{\begingroup\mn@urlcharsother \@ifnextchar [ {\mn@doi@}
  {\mn@doi@[]}}
\def\mn@doi@[#1]#2{\def\@tempa{#1}\ifx\@tempa\@empty \href
  {http://dx.doi.org/#2} {doi:#2}\else \href {http://dx.doi.org/#2} {#1}\fi
  \endgroup}
\def\mn@eprint#1#2{\mn@eprint@#1:#2::\@nil}
\def\mn@eprint@arXiv#1{\href {http://arxiv.org/abs/#1} {{\tt arXiv:#1}}}
\def\mn@eprint@dblp#1{\href {http://dblp.uni-trier.de/rec/bibtex/#1.xml}
  {dblp:#1}}
\def\mn@eprint@#1:#2:#3:#4\@nil{\def\@tempa {#1}\def\@tempb {#2}\def\@tempc
  {#3}\ifx \@tempc \@empty \let \@tempc \@tempb \let \@tempb \@tempa \fi \ifx
  \@tempb \@empty \def\@tempb {arXiv}\fi \@ifundefined
  {mn@eprint@\@tempb}{\@tempb:\@tempc}{\expandafter \expandafter \csname
  mn@eprint@\@tempb\endcsname \expandafter{\@tempc}}}

\bibitem[\protect\citeauthoryear{{Abraham} \& {van Dokkum}}{{Abraham} \& {van
  Dokkum}}{2014}]{Abraham2014}
{Abraham} R.~G.,  {van Dokkum} P.~G.,  2014, \mn@doi [\pasp] {10.1086/674875},
  \href {https://ui.adsabs.harvard.edu/abs/2014PASP..126...55A} {126, 55}

\bibitem[\protect\citeauthoryear{{Amorisco} \& {Loeb}}{{Amorisco} \&
  {Loeb}}{2016}]{Amorisco2016}
{Amorisco} N.~C.,  {Loeb} A.,  2016, \mn@doi [\mnras] {10.1093/mnrasl/slw055},
  \href {https://ui.adsabs.harvard.edu/abs/2016MNRAS.459L..51A} {459, L51}

\bibitem[\protect\citeauthoryear{{Amorisco}, {Monachesi}, {Agnello}  \&
  {White}}{{Amorisco} et~al.}{2018}]{amorisco2018}
{Amorisco} N.~C.,  {Monachesi} A.,  {Agnello} A.,   {White} S.~D.~M.,  2018,
  \mn@doi [\mnras] {10.1093/mnras/sty116}, \href
  {https://ui.adsabs.harvard.edu/abs/2018MNRAS.475.4235A} {475, 4235}

\bibitem[\protect\citeauthoryear{{Beasley}, {Romanowsky}, {Pota}, {Navarro},
  {Martinez Delgado}, {Neyer}  \& {Deich}}{{Beasley}
  et~al.}{2016}]{Beasley2016}
{Beasley} M.~A.,  {Romanowsky} A.~J.,  {Pota} V.,  {Navarro} I.~M.,  {Martinez
  Delgado} D.,  {Neyer} F.,   {Deich} A.~L.,  2016, \mn@doi [\apjl]
  {10.3847/2041-8205/819/2/L20}, \href
  {https://ui.adsabs.harvard.edu/abs/2016ApJ...819L..20B} {819, L20}

\bibitem[\protect\citeauthoryear{{Benavides} et~al.,}{{Benavides}
  et~al.}{2021}]{Benavides2021}
{Benavides} J.~A.,  et~al., 2021, \mn@doi [Nature Astronomy]
  {10.1038/s41550-021-01458-1}, \href
  {https://ui.adsabs.harvard.edu/abs/2021NatAs...5.1255B} {5, 1255}

\bibitem[\protect\citeauthoryear{{Benavides}, {Sales}, {Abadi}, {Marinacci},
  {Vogelsberger}  \& {Hernquist}}{{Benavides} et~al.}{2023}]{Benavides2023}
{Benavides} J.~A.,  {Sales} L.~V.,  {Abadi} M.~G.,  {Marinacci} F.,
  {Vogelsberger} M.,   {Hernquist} L.,  2023, \mn@doi [\mnras]
  {10.1093/mnras/stad1053}, \href
  {https://ui.adsabs.harvard.edu/abs/2023MNRAS.522.1033B} {522, 1033}

\bibitem[\protect\citeauthoryear{{Binggeli}, {Sandage}  \&
  {Tammann}}{{Binggeli} et~al.}{1985}]{Binggeli1985}
{Binggeli} B.,  {Sandage} A.,   {Tammann} G.~A.,  1985, \mn@doi [\aj]
  {10.1086/113874}, \href
  {https://ui.adsabs.harvard.edu/abs/1985AJ.....90.1681B} {90, 1681}

\bibitem[\protect\citeauthoryear{Binney \& Tremaine}{Binney \&
  Tremaine}{2008}]{bandt2008}
Binney J.,  Tremaine S.,  2008, Galactic Dynamics: Second Edition, rev -
  revised, 2 edn.
Princeton University Press, \url {http://www.jstor.org/stable/j.ctvc778ff}

\bibitem[\protect\citeauthoryear{{Bothun}, {Impey}  \& {Malin}}{{Bothun}
  et~al.}{1991}]{Bothun1991}
{Bothun} G.~D.,  {Impey} C.~D.,   {Malin} D.~F.,  1991, \mn@doi [\apj]
  {10.1086/170290}, \href
  {https://ui.adsabs.harvard.edu/abs/1991ApJ...376..404B} {376, 404}

\bibitem[\protect\citeauthoryear{{Carleton}, {Errani}, {Cooper}, {Kaplinghat},
  {Pe{\~n}arrubia}  \& {Guo}}{{Carleton} et~al.}{2019}]{Carleton2019}
{Carleton} T.,  {Errani} R.,  {Cooper} M.,  {Kaplinghat} M.,  {Pe{\~n}arrubia}
  J.,   {Guo} Y.,  2019, \mn@doi [\mnras] {10.1093/mnras/stz383}, \href
  {https://ui.adsabs.harvard.edu/abs/2019MNRAS.485..382C} {485, 382}

\bibitem[\protect\citeauthoryear{{Chan}, {Kere{\v{s}}}, {Wetzel}, {Hopkins},
  {Faucher-Gigu{\`e}re}, {El-Badry}, {Garrison-Kimmel}  \&
  {Boylan-Kolchin}}{{Chan} et~al.}{2018}]{Chan2018}
{Chan} T.~K.,  {Kere{\v{s}}} D.,  {Wetzel} A.,  {Hopkins} P.~F.,
  {Faucher-Gigu{\`e}re} C.~A.,  {El-Badry} K.,  {Garrison-Kimmel} S.,
  {Boylan-Kolchin} M.,  2018, \mn@doi [\mnras] {10.1093/mnras/sty1153}, \href
  {https://ui.adsabs.harvard.edu/abs/2018MNRAS.478..906C} {478, 906}

\bibitem[\protect\citeauthoryear{{Chilingarian}, {Afanasiev}, {Grishin},
  {Fabricant}  \& {Moran}}{{Chilingarian} et~al.}{2019}]{Chilingarian2019}
{Chilingarian} I.~V.,  {Afanasiev} A.~V.,  {Grishin} K.~A.,  {Fabricant} D.,
  {Moran} S.,  2019, \mn@doi [\apj] {10.3847/1538-4357/ab4205}, \href
  {https://ui.adsabs.harvard.edu/abs/2019ApJ...884...79C} {884, 79}

\bibitem[\protect\citeauthoryear{{Dalcanton}, {Spergel}, {Gunn}, {Schmidt}  \&
  {Schneider}}{{Dalcanton} et~al.}{1997}]{Dalcanton1997}
{Dalcanton} J.~J.,  {Spergel} D.~N.,  {Gunn} J.~E.,  {Schmidt} M.,
  {Schneider} D.~P.,  1997, \mn@doi [\aj] {10.1086/118499}, \href
  {https://ui.adsabs.harvard.edu/abs/1997AJ....114..635D} {114, 635}

\bibitem[\protect\citeauthoryear{{Danieli}, {van Dokkum}, {Conroy}, {Abraham}
  \& {Romanowsky}}{{Danieli} et~al.}{2019}]{Danieli2019}
{Danieli} S.,  {van Dokkum} P.,  {Conroy} C.,  {Abraham} R.,   {Romanowsky}
  A.~J.,  2019, \mn@doi [\apjl] {10.3847/2041-8213/ab0e8c}, \href
  {https://ui.adsabs.harvard.edu/abs/2019ApJ...874L..12D} {874, L12}

\bibitem[\protect\citeauthoryear{{Danieli} et~al.,}{{Danieli}
  et~al.}{2022}]{Danieli2022}
{Danieli} S.,  et~al., 2022, \mn@doi [\apjl] {10.3847/2041-8213/ac590a}, \href
  {https://ui.adsabs.harvard.edu/abs/2022ApJ...927L..28D} {927, L28}

\bibitem[\protect\citeauthoryear{{Davis}, {Efstathiou}, {Frenk}  \&
  {White}}{{Davis} et~al.}{1985}]{davis1985}
{Davis} M.,  {Efstathiou} G.,  {Frenk} C.~S.,   {White} S.~D.~M.,  1985,
  \mn@doi [\apj] {10.1086/163168}, \href
  {https://ui.adsabs.harvard.edu/abs/1985ApJ...292..371D} {292, 371}

\bibitem[\protect\citeauthoryear{{Di Cintio}, {Brook}, {Dutton}, {Macci{\`o}},
  {Obreja}  \& {Dekel}}{{Di Cintio} et~al.}{2017}]{DiCintio2017}
{Di Cintio} A.,  {Brook} C.~B.,  {Dutton} A.~A.,  {Macci{\`o}} A.~V.,  {Obreja}
  A.,   {Dekel} A.,  2017, \mn@doi [\mnras] {10.1093/mnrasl/slw210}, \href
  {https://ui.adsabs.harvard.edu/abs/2017MNRAS.466L...1D} {466, L1}

\bibitem[\protect\citeauthoryear{{Dolag}, {Borgani}, {Murante}  \&
  {Springel}}{{Dolag} et~al.}{2009}]{dolag2009}
{Dolag} K.,  {Borgani} S.,  {Murante} G.,   {Springel} V.,  2009, \mn@doi
  [\mnras] {10.1111/j.1365-2966.2009.15034.x}, \href
  {https://ui.adsabs.harvard.edu/abs/2009MNRAS.399..497D} {399, 497}

\bibitem[\protect\citeauthoryear{{Doppel}, {Sales}, {Navarro}, {Abadi}, {Peng},
  {Toloba}  \& {Ramos-Almendares}}{{Doppel} et~al.}{2021}]{doppel2021}
{Doppel} J.~E.,  {Sales} L.~V.,  {Navarro} J.~F.,  {Abadi} M.~G.,  {Peng}
  E.~W.,  {Toloba} E.,   {Ramos-Almendares} F.,  2021, \mn@doi [\mnras]
  {10.1093/mnras/staa3915}, \href
  {https://ui.adsabs.harvard.edu/abs/2021MNRAS.502.1661D} {502, 1661}

\bibitem[\protect\citeauthoryear{{Doppel} et~al.,}{{Doppel}
  et~al.}{2023}]{Doppel2023}
{Doppel} J.~E.,  et~al., 2023, \mn@doi [\mnras] {10.1093/mnras/stac2818}, \href
  {https://ui.adsabs.harvard.edu/abs/2023MNRAS.518.2453D} {518, 2453}

\bibitem[\protect\citeauthoryear{{Durrell} et~al.,}{{Durrell}
  et~al.}{2014}]{durell2014}
{Durrell} P.~R.,  et~al., 2014, \mn@doi [\apj] {10.1088/0004-637X/794/2/103},
  \href {https://ui.adsabs.harvard.edu/abs/2014ApJ...794..103D} {794, 103}

\bibitem[\protect\citeauthoryear{Ferré-Mateu et~al.,}{Ferré-Mateu
  et~al.}{2018}]{FerreMateu2018}
Ferré-Mateu A.,  et~al., 2018, \mn@doi [Monthly Notices of the Royal
  Astronomical Society] {10.1093/mnras/sty1597}, 479, 4891

\bibitem[\protect\citeauthoryear{Forbes, Read, Gieles  \& Collins}{Forbes
  et~al.}{2018}]{Forbes2018}
Forbes D.~A.,  Read J.~I.,  Gieles M.,   Collins M. L.~M.,  2018, \mn@doi
  [Monthly Notices of the Royal Astronomical Society] {10.1093/mnras/sty2584},
  481, 5592

\bibitem[\protect\citeauthoryear{{Forbes}, {Alabi}, {Romanowsky}, {Brodie}  \&
  {Arimoto}}{{Forbes} et~al.}{2020}]{Forbes2020}
{Forbes} D.~A.,  {Alabi} A.,  {Romanowsky} A.~J.,  {Brodie} J.~P.,   {Arimoto}
  N.,  2020, \mn@doi [\mnras] {10.1093/mnras/staa180}, \href
  {https://ui.adsabs.harvard.edu/abs/2020MNRAS.492.4874F} {492, 4874}

\bibitem[\protect\citeauthoryear{{Gannon} et~al.,}{{Gannon}
  et~al.}{2021}]{Gannon2021}
{Gannon} J.~S.,  et~al., 2021, \mn@doi [\mnras] {10.1093/mnras/stab277}, \href
  {https://ui.adsabs.harvard.edu/abs/2021MNRAS.502.3144G} {502, 3144}

\bibitem[\protect\citeauthoryear{{Gannon} et~al.,}{{Gannon}
  et~al.}{2022}]{Gannon2022}
{Gannon} J.~S.,  et~al., 2022, \mn@doi [\mnras] {10.1093/mnras/stab3297}, \href
  {https://ui.adsabs.harvard.edu/abs/2022MNRAS.510..946G} {510, 946}

\bibitem[\protect\citeauthoryear{{Gannon}, {Forbes}, {Brodie}, {Romanowsky},
  {Couch}  \& {Ferr{\'e}-Mateu}}{{Gannon} et~al.}{2023}]{Gannon2023}
{Gannon} J.~S.,  {Forbes} D.~A.,  {Brodie} J.~P.,  {Romanowsky} A.~J.,  {Couch}
  W.~J.,   {Ferr{\'e}-Mateu} A.,  2023, \mn@doi [\mnras]
  {10.1093/mnras/stac3264}, \href
  {https://ui.adsabs.harvard.edu/abs/2023MNRAS.518.3653G} {518, 3653}

\bibitem[\protect\citeauthoryear{{Harris} \& {van den Bergh}}{{Harris} \& {van
  den Bergh}}{1981}]{harris1981}
{Harris} W.~E.,  {van den Bergh} S.,  1981, \mn@doi [\aj] {10.1086/113047},
  \href {https://ui.adsabs.harvard.edu/abs/1981AJ.....86.1627H} {86, 1627}

\bibitem[\protect\citeauthoryear{Harris, Harris  \& Hudson}{Harris
  et~al.}{2015}]{Harris2015}
Harris W.~E.,  Harris G.~L.,   Hudson M.~J.,  2015, \mn@doi [The Astrophysical
  Journal] {10.1088/0004-637x/806/1/36}, 806, 36

\bibitem[\protect\citeauthoryear{{Hernquist}}{{Hernquist}}{1990}]{hernquist1990}
{Hernquist} L.,  1990, \mn@doi [\apj] {10.1086/168845}, \href
  {https://ui.adsabs.harvard.edu/abs/1990ApJ...356..359H} {356, 359}

\bibitem[\protect\citeauthoryear{{Impey}, {Bothun}  \& {Malin}}{{Impey}
  et~al.}{1988}]{Impey1988}
{Impey} C.,  {Bothun} G.,   {Malin} D.,  1988, \mn@doi [\apj] {10.1086/166500},
  \href {https://ui.adsabs.harvard.edu/abs/1988ApJ...330..634I} {330, 634}

\bibitem[\protect\citeauthoryear{{Janssens} et~al.,}{{Janssens}
  et~al.}{2022}]{Janssens2022}
{Janssens} S.~R.,  et~al., 2022, \mn@doi [\mnras] {10.1093/mnras/stac2717},
  \href {https://ui.adsabs.harvard.edu/abs/2022MNRAS.517..858J} {517, 858}

\bibitem[\protect\citeauthoryear{{Jiang}, {Dekel}, {Freundlich}, {Romanowsky},
  {Dutton}, {Macci{\`o}}  \& {Di Cintio}}{{Jiang} et~al.}{2019}]{Jiang2019}
{Jiang} F.,  {Dekel} A.,  {Freundlich} J.,  {Romanowsky} A.~J.,  {Dutton}
  A.~A.,  {Macci{\`o}} A.~V.,   {Di Cintio} A.,  2019, \mn@doi [\mnras]
  {10.1093/mnras/stz1499}, \href
  {https://ui.adsabs.harvard.edu/abs/2019MNRAS.487.5272J} {487, 5272}

\bibitem[\protect\citeauthoryear{{Jones} et~al.,}{{Jones}
  et~al.}{2023}]{Jones2023}
{Jones} M.~G.,  et~al., 2023, \mn@doi [\apjl] {10.3847/2041-8213/acaaab}, \href
  {https://ui.adsabs.harvard.edu/abs/2023ApJ...942L...5J} {942, L5}

\bibitem[\protect\citeauthoryear{{Jord{\'a}n} et~al.,}{{Jord{\'a}n}
  et~al.}{2007}]{jordan2007}
{Jord{\'a}n} A.,  et~al., 2007, \mn@doi [\apjs] {10.1086/516840}, \href
  {https://ui.adsabs.harvard.edu/abs/2007ApJS..171..101J} {171, 101}

\bibitem[\protect\citeauthoryear{{Junais} et~al.,}{{Junais}
  et~al.}{2021}]{Junais2021}
{Junais} et~al., 2021, \mn@doi [\aap] {10.1051/0004-6361/202040185}, \href
  {https://ui.adsabs.harvard.edu/abs/2021A&A...650A..99J} {650, A99}

\bibitem[\protect\citeauthoryear{{Karachentsev} \& {Nasonova}}{{Karachentsev}
  \& {Nasonova}}{2010}]{karachentsev2010}
{Karachentsev} I.~D.,  {Nasonova} O.~G.,  2010, \mn@doi [\mnras]
  {10.1111/j.1365-2966.2010.16501.x}, \href
  {https://ui.adsabs.harvard.edu/abs/2010MNRAS.405.1075K} {405, 1075}

\bibitem[\protect\citeauthoryear{{Koda}, {Yagi}, {Yamanoi}  \&
  {Komiyama}}{{Koda} et~al.}{2015}]{Koda2015}
{Koda} J.,  {Yagi} M.,  {Yamanoi} H.,   {Komiyama} Y.,  2015, \mn@doi [\apjl]
  {10.1088/2041-8205/807/1/L2}, \href
  {https://ui.adsabs.harvard.edu/abs/2015ApJ...807L...2K} {807, L2}

\bibitem[\protect\citeauthoryear{{Kong}, {Kaplinghat}, {Yu}, {Fraternali}  \&
  {Mancera Pi{\~n}a}}{{Kong} et~al.}{2022}]{Kong2022}
{Kong} D.,  {Kaplinghat} M.,  {Yu} H.-B.,  {Fraternali} F.,   {Mancera
  Pi{\~n}a} P.~E.,  2022, \mn@doi [\apj] {10.3847/1538-4357/ac8875}, \href
  {https://ui.adsabs.harvard.edu/abs/2022ApJ...936..166K} {936, 166}

\bibitem[\protect\citeauthoryear{{La Marca} et~al.,}{{La Marca}
  et~al.}{2022}]{LaMarca2022}
{La Marca} A.,  et~al., 2022, arXiv e-prints, \href
  {https://ui.adsabs.harvard.edu/abs/2022arXiv220607385L} {p. arXiv:2206.07385}

\bibitem[\protect\citeauthoryear{{Lah{\'e}n}, {Naab}, {Johansson}, {Elmegreen},
  {Hu}, {Walch}, {Steinwandel}  \& {Moster}}{{Lah{\'e}n}
  et~al.}{2020}]{Lahen2020}
{Lah{\'e}n} N.,  {Naab} T.,  {Johansson} P.~H.,  {Elmegreen} B.,  {Hu} C.-Y.,
  {Walch} S.,  {Steinwandel} U.~P.,   {Moster} B.~P.,  2020, \mn@doi [\apj]
  {10.3847/1538-4357/ab7190}, \href
  {https://ui.adsabs.harvard.edu/abs/2020ApJ...891....2L} {891, 2}

\bibitem[\protect\citeauthoryear{{Lee}, {Park}  \& {Hwang}}{{Lee}
  et~al.}{2010}]{lee2010}
{Lee} M.~G.,  {Park} H.~S.,   {Hwang} H.~S.,  2010, \mn@doi [Science]
  {10.1126/science.1186496}, \href
  {https://ui.adsabs.harvard.edu/abs/2010Sci...328..334L} {328, 334}

\bibitem[\protect\citeauthoryear{{Lee}, {Kang}, {Lee}  \& {Jang}}{{Lee}
  et~al.}{2017}]{Lee2017}
{Lee} M.~G.,  {Kang} J.,  {Lee} J.~H.,   {Jang} I.~S.,  2017, \mn@doi [\apj]
  {10.3847/1538-4357/aa78fb}, \href
  {https://ui.adsabs.harvard.edu/abs/2017ApJ...844..157L} {844, 157}

\bibitem[\protect\citeauthoryear{{Lee}, {Kang}, {Lee}  \& {Jang}}{{Lee}
  et~al.}{2020}]{Lee2020}
{Lee} J.~H.,  {Kang} J.,  {Lee} M.~G.,   {Jang} I.~S.,  2020, \mn@doi [\apj]
  {10.3847/1538-4357/ab8632}, \href
  {https://ui.adsabs.harvard.edu/abs/2020ApJ...894...75L} {894, 75}

\bibitem[\protect\citeauthoryear{{Leisman} et~al.,}{{Leisman}
  et~al.}{2017}]{Leisman2017}
{Leisman} L.,  et~al., 2017, \mn@doi [\apj] {10.3847/1538-4357/aa7575}, \href
  {https://ui.adsabs.harvard.edu/abs/2017ApJ...842..133L} {842, 133}

\bibitem[\protect\citeauthoryear{{Lim}, {Peng}, {C{\^o}t{\'e}}, {Sales}, {den
  Brok}, {Blakeslee}  \& {Guhathakurta}}{{Lim} et~al.}{2018}]{Lim2018}
{Lim} S.,  {Peng} E.~W.,  {C{\^o}t{\'e}} P.,  {Sales} L.~V.,  {den Brok} M.,
  {Blakeslee} J.~P.,   {Guhathakurta} P.,  2018, \mn@doi [\apj]
  {10.3847/1538-4357/aacb81}, \href
  {https://ui.adsabs.harvard.edu/abs/2018ApJ...862...82L} {862, 82}

\bibitem[\protect\citeauthoryear{{Lim} et~al.,}{{Lim} et~al.}{2020}]{Lim2020}
{Lim} S.,  et~al., 2020, \mn@doi [\apj] {10.3847/1538-4357/aba433}, \href
  {https://ui.adsabs.harvard.edu/abs/2020ApJ...899...69L} {899, 69}

\bibitem[\protect\citeauthoryear{{Macci{\`o}}, {Prats}, {Dixon}, {Buck},
  {Waterval}, {Arora}, {Courteau}  \& {Kang}}{{Macci{\`o}}
  et~al.}{2021}]{Maccio2021}
{Macci{\`o}} A.~V.,  {Prats} D.~H.,  {Dixon} K.~L.,  {Buck} T.,  {Waterval} S.,
   {Arora} N.,  {Courteau} S.,   {Kang} X.,  2021, \mn@doi [\mnras]
  {10.1093/mnras/staa3716}, \href
  {https://ui.adsabs.harvard.edu/abs/2021MNRAS.501..693M} {501, 693}

\bibitem[\protect\citeauthoryear{{Mancera Pi{\~n}a} et~al.,}{{Mancera Pi{\~n}a}
  et~al.}{2020}]{ManceraPina2020}
{Mancera Pi{\~n}a} P.~E.,  et~al., 2020, \mn@doi [\mnras]
  {10.1093/mnras/staa1256}, \href
  {https://ui.adsabs.harvard.edu/abs/2020MNRAS.495.3636M} {495, 3636}

\bibitem[\protect\citeauthoryear{{Marinacci} et~al.,}{{Marinacci}
  et~al.}{2018}]{marinacci2018}
{Marinacci} F.,  et~al., 2018, \mn@doi [\mnras] {10.1093/mnras/sty2206}, \href
  {https://ui.adsabs.harvard.edu/abs/2018MNRAS.480.5113M} {480, 5113}

\bibitem[\protect\citeauthoryear{{Marleau} et~al.,}{{Marleau}
  et~al.}{2021}]{Marleau2021}
{Marleau} F.~R.,  et~al., 2021, \mn@doi [\aap] {10.1051/0004-6361/202141432},
  \href {https://ui.adsabs.harvard.edu/abs/2021A&A...654A.105M} {654, A105}

\bibitem[\protect\citeauthoryear{{Mart{\'\i}n-Navarro}
  et~al.,}{{Mart{\'\i}n-Navarro} et~al.}{2019}]{MartinNavarro2019}
{Mart{\'\i}n-Navarro} I.,  et~al., 2019, \mn@doi [\mnras]
  {10.1093/mnras/stz252}, \href
  {https://ui.adsabs.harvard.edu/abs/2019MNRAS.484.3425M} {484, 3425}

\bibitem[\protect\citeauthoryear{{Mart{\'\i}nez-Delgado}
  et~al.,}{{Mart{\'\i}nez-Delgado} et~al.}{2016}]{MartinexDelgado2016}
{Mart{\'\i}nez-Delgado} D.,  et~al., 2016, \mn@doi [\aj]
  {10.3847/0004-6256/151/4/96}, \href
  {https://ui.adsabs.harvard.edu/abs/2016AJ....151...96M} {151, 96}

\bibitem[\protect\citeauthoryear{{Mihos} et~al.,}{{Mihos}
  et~al.}{2015}]{Mihos2015}
{Mihos} J.~C.,  et~al., 2015, \mn@doi [\apjl] {10.1088/2041-8205/809/2/L21},
  \href {https://ui.adsabs.harvard.edu/abs/2015ApJ...809L..21M} {809, L21}

\bibitem[\protect\citeauthoryear{{Moreno} et~al.,}{{Moreno}
  et~al.}{2022}]{Moreno2022}
{Moreno} J.,  et~al., 2022, \mn@doi [Nature Astronomy]
  {10.1038/s41550-021-01598-4}, \href
  {https://ui.adsabs.harvard.edu/abs/2022NatAs...6..496M} {6, 496}

\bibitem[\protect\citeauthoryear{{M{\"u}ller} et~al.,}{{M{\"u}ller}
  et~al.}{2021}]{Muller2021}
{M{\"u}ller} O.,  et~al., 2021, \mn@doi [\apj] {10.3847/1538-4357/ac2831},
  \href {https://ui.adsabs.harvard.edu/abs/2021ApJ...923....9M} {923, 9}

\bibitem[\protect\citeauthoryear{{Naiman} et~al.,}{{Naiman}
  et~al.}{2018}]{naiman2018}
{Naiman} J.~P.,  et~al., 2018, \mn@doi [\mnras] {10.1093/mnras/sty618}, \href
  {https://ui.adsabs.harvard.edu/abs/2018MNRAS.477.1206N} {477, 1206}

\bibitem[\protect\citeauthoryear{{Navarro}, {Frenk}  \& {White}}{{Navarro}
  et~al.}{1996}]{navarro1996}
{Navarro} J.~F.,  {Frenk} C.~S.,   {White} S. D.~M.,  1996, \mn@doi [\apj]
  {10.1086/177173}, \href
  {https://ui.adsabs.harvard.edu/abs/1996ApJ...462..563N} {462, 563}

\bibitem[\protect\citeauthoryear{{Navarro}, {Frenk}  \& {White}}{{Navarro}
  et~al.}{1997}]{navarro1997}
{Navarro} J.~F.,  {Frenk} C.~S.,   {White} S. D.~M.,  1997, \mn@doi [\apj]
  {10.1086/304888}, \href
  {https://ui.adsabs.harvard.edu/abs/1997ApJ...490..493N} {490, 493}

\bibitem[\protect\citeauthoryear{{Nelson} et~al.,}{{Nelson}
  et~al.}{2018}]{nelson2018}
{Nelson} D.,  et~al., 2018, \mn@doi [\mnras] {10.1093/mnras/stx3040}, \href
  {https://ui.adsabs.harvard.edu/abs/2018MNRAS.475..624N} {475, 624}

\bibitem[\protect\citeauthoryear{{Nelson} et~al.,}{{Nelson}
  et~al.}{2019a}]{nelson2019dr}
{Nelson} D.,  et~al., 2019a, \mn@doi [Computational Astrophysics and Cosmology]
  {10.1186/s40668-019-0028-x}, \href
  {https://ui.adsabs.harvard.edu/abs/2019ComAC...6....2N} {6, 2}

\bibitem[\protect\citeauthoryear{{Nelson} et~al.,}{{Nelson}
  et~al.}{2019b}]{nelson2019}
{Nelson} D.,  et~al., 2019b, \mn@doi [\mnras] {10.1093/mnras/stz2306}, \href
  {https://ui.adsabs.harvard.edu/abs/2019MNRAS.490.3234N} {490, 3234}

\bibitem[\protect\citeauthoryear{{Papastergis}, {Adams}  \&
  {Romanowsky}}{{Papastergis} et~al.}{2017}]{Papastergis2017}
{Papastergis} E.,  {Adams} E.~A.~K.,   {Romanowsky} A.~J.,  2017, \mn@doi
  [\aap] {10.1051/0004-6361/201730795}, \href
  {https://ui.adsabs.harvard.edu/abs/2017A&A...601L..10P} {601, L10}

\bibitem[\protect\citeauthoryear{{Peng} \& {Lim}}{{Peng} \&
  {Lim}}{2016}]{Peng2016}
{Peng} E.~W.,  {Lim} S.,  2016, \mn@doi [\apjl] {10.3847/2041-8205/822/2/L31},
  \href {https://ui.adsabs.harvard.edu/abs/2016ApJ...822L..31P} {822, L31}

\bibitem[\protect\citeauthoryear{Peng et~al.,}{Peng et~al.}{2008}]{Peng2008}
Peng E.~W.,  et~al., 2008, \mn@doi [The Astrophysical Journal]
  {10.1086/587951}, 681, 197

\bibitem[\protect\citeauthoryear{{Pillepich} et~al.,}{{Pillepich}
  et~al.}{2018a}]{pillepich2018}
{Pillepich} A.,  et~al., 2018a, \mn@doi [\mnras] {10.1093/mnras/stx3112}, \href
  {https://ui.adsabs.harvard.edu/abs/2018MNRAS.475..648P} {475, 648}

\bibitem[\protect\citeauthoryear{{Pillepich} et~al.,}{{Pillepich}
  et~al.}{2018b}]{pillepich2018b}
{Pillepich} A.,  et~al., 2018b, \mn@doi [\mnras] {10.1093/mnras/stx3112}, \href
  {https://ui.adsabs.harvard.edu/abs/2018MNRAS.475..648P} {475, 648}

\bibitem[\protect\citeauthoryear{{Pillepich} et~al.,}{{Pillepich}
  et~al.}{2019}]{pillepich2019}
{Pillepich} A.,  et~al., 2019, \mn@doi [\mnras] {10.1093/mnras/stz2338}, \href
  {https://ui.adsabs.harvard.edu/abs/2019MNRAS.490.3196P} {490, 3196}

\bibitem[\protect\citeauthoryear{{Planck Collaboration} et~al.,}{{Planck
  Collaboration} et~al.}{2016}]{planckcollab2016}
{Planck Collaboration} et~al., 2016, \mn@doi [A\&A]
  {10.1051/0004-6361/201525830}, 594, A13

\bibitem[\protect\citeauthoryear{Prole et~al.,}{Prole et~al.}{2019}]{Prole2019}
Prole D.~J.,  et~al., 2019, \mn@doi [Monthly Notices of the Royal Astronomical
  Society] {10.1093/mnras/stz326}, 484, 4865

\bibitem[\protect\citeauthoryear{{Ramos-Almendares}, {Sales}, {Abadi},
  {Doppel}, {Muriel}  \& {Peng}}{{Ramos-Almendares}
  et~al.}{2020}]{ramosalmendares2020}
{Ramos-Almendares} F.,  {Sales} L.~V.,  {Abadi} M.~G.,  {Doppel} J.~E.,
  {Muriel} H.,   {Peng} E.~W.,  2020, \mn@doi [\mnras] {10.1093/mnras/staa551},
  \href {https://ui.adsabs.harvard.edu/abs/2020MNRAS.493.5357R} {493, 5357}

\bibitem[\protect\citeauthoryear{{Reaves}}{{Reaves}}{1983}]{Reaves1983}
{Reaves} G.,  1983, \mn@doi [\apjs] {10.1086/190895}, \href
  {https://ui.adsabs.harvard.edu/abs/1983ApJS...53..375R} {53, 375}

\bibitem[\protect\citeauthoryear{Rodriguez-Gomez et~al.,}{Rodriguez-Gomez
  et~al.}{2015}]{rodriguez-gomez2015}
Rodriguez-Gomez V.,  et~al., 2015, \mn@doi [Monthly Notices of the Royal
  Astronomical Society] {10.1093/mnras/stv264}, 449, 49

\bibitem[\protect\citeauthoryear{{Rom{\'a}n} \& {Trujillo}}{{Rom{\'a}n} \&
  {Trujillo}}{2017}]{Roman2017}
{Rom{\'a}n} J.,  {Trujillo} I.,  2017, \mn@doi [\mnras] {10.1093/mnras/stx694},
  \href {https://ui.adsabs.harvard.edu/abs/2017MNRAS.468.4039R} {468, 4039}

\bibitem[\protect\citeauthoryear{{Rom{\'a}n}, {Beasley}, {Ruiz-Lara}  \&
  {Valls-Gabaud}}{{Rom{\'a}n} et~al.}{2019}]{Roman2019}
{Rom{\'a}n} J.,  {Beasley} M.~A.,  {Ruiz-Lara} T.,   {Valls-Gabaud} D.,  2019,
  \mn@doi [\mnras] {10.1093/mnras/stz835}, \href
  {https://ui.adsabs.harvard.edu/abs/2019MNRAS.486..823R} {486, 823}

\bibitem[\protect\citeauthoryear{{Rong}, {Guo}, {Gao}, {Liao}, {Xie}, {Puzia},
  {Sun}  \& {Pan}}{{Rong} et~al.}{2017}]{Rong2017}
{Rong} Y.,  {Guo} Q.,  {Gao} L.,  {Liao} S.,  {Xie} L.,  {Puzia} T.~H.,  {Sun}
  S.,   {Pan} J.,  2017, \mn@doi [\mnras] {10.1093/mnras/stx1440}, \href
  {https://ui.adsabs.harvard.edu/abs/2017MNRAS.470.4231R} {470, 4231}

\bibitem[\protect\citeauthoryear{{Rong}, {Mancera Pi{\~n}a}, {Tempel}, {Puzia}
  \& {De Rijcke}}{{Rong} et~al.}{2020a}]{Rong2020b}
{Rong} Y.,  {Mancera Pi{\~n}a} P.~E.,  {Tempel} E.,  {Puzia} T.~H.,   {De
  Rijcke} S.,  2020a, \mn@doi [\mnras] {10.1093/mnrasl/slaa129}, \href
  {https://ui.adsabs.harvard.edu/abs/2020MNRAS.498L..72R} {498, L72}

\bibitem[\protect\citeauthoryear{{Rong}, {Zhu}, {Johnston}, {Zhang}, {Cao},
  {Puzia}  \& {Galaz}}{{Rong} et~al.}{2020b}]{Rong2020a}
{Rong} Y.,  {Zhu} K.,  {Johnston} E.~J.,  {Zhang} H.-X.,  {Cao} T.,  {Puzia}
  T.~H.,   {Galaz} G.,  2020b, \mn@doi [\apjl] {10.3847/2041-8213/aba8aa},
  \href {https://ui.adsabs.harvard.edu/abs/2020ApJ...899L..12R} {899, L12}

\bibitem[\protect\citeauthoryear{{Saifollahi}, {Trujillo}, {Beasley},
  {Peletier}  \& {Knapen}}{{Saifollahi} et~al.}{2021}]{Saifollahi2021}
{Saifollahi} T.,  {Trujillo} I.,  {Beasley} M.~A.,  {Peletier} R.~F.,
  {Knapen} J.~H.,  2021, \mn@doi [\mnras] {10.1093/mnras/staa3016}, \href
  {https://ui.adsabs.harvard.edu/abs/2021MNRAS.502.5921S} {502, 5921}

\bibitem[\protect\citeauthoryear{{Sales}, {Navarro}, {Pe{\~n}afiel}, {Peng},
  {Lim}  \& {Hernquist}}{{Sales} et~al.}{2020}]{Sales2020}
{Sales} L.~V.,  {Navarro} J.~F.,  {Pe{\~n}afiel} L.,  {Peng} E.~W.,  {Lim} S.,
   {Hernquist} L.,  2020, \mn@doi [\mnras] {10.1093/mnras/staa854}, \href
  {https://ui.adsabs.harvard.edu/abs/2020MNRAS.494.1848S} {494, 1848}

\bibitem[\protect\citeauthoryear{{Somalwar}, {Greene}, {Greco}, {Huang},
  {Beaton}, {Goulding}  \& {Lancaster}}{{Somalwar} et~al.}{2020}]{Somalwar2020}
{Somalwar} J.~J.,  {Greene} J.~E.,  {Greco} J.~P.,  {Huang} S.,  {Beaton}
  R.~L.,  {Goulding} A.~D.,   {Lancaster} L.,  2020, \mn@doi [\apj]
  {10.3847/1538-4357/abb1b2}, \href
  {https://ui.adsabs.harvard.edu/abs/2020ApJ...902...45S} {902, 45}

\bibitem[\protect\citeauthoryear{{Somerville} et~al.,}{{Somerville}
  et~al.}{2018}]{Somerville2018}
{Somerville} R.~S.,  et~al., 2018, \mn@doi [\mnras] {10.1093/mnras/stx2040},
  \href {https://ui.adsabs.harvard.edu/abs/2018MNRAS.473.2714S} {473, 2714}

\bibitem[\protect\citeauthoryear{{Springel}, {White}, {Tormen}  \&
  {Kauffmann}}{{Springel} et~al.}{2001}]{springel2001}
{Springel} V.,  {White} S. D.~M.,  {Tormen} G.,   {Kauffmann} G.,  2001,
  \mn@doi [\mnras] {10.1046/j.1365-8711.2001.04912.x}, \href
  {https://ui.adsabs.harvard.edu/abs/2001MNRAS.328..726S} {328, 726}

\bibitem[\protect\citeauthoryear{{Springel} et~al.,}{{Springel}
  et~al.}{2018}]{springel2018}
{Springel} V.,  et~al., 2018, \mn@doi [\mnras] {10.1093/mnras/stx3304}, \href
  {https://ui.adsabs.harvard.edu/abs/2018MNRAS.475..676S} {475, 676}

\bibitem[\protect\citeauthoryear{{Toloba} et~al.,}{{Toloba}
  et~al.}{2018}]{Toloba2018}
{Toloba} E.,  et~al., 2018, \mn@doi [\apjl] {10.3847/2041-8213/aab603}, \href
  {https://ui.adsabs.harvard.edu/abs/2018ApJ...856L..31T} {856, L31}

\bibitem[\protect\citeauthoryear{{Toloba} et~al.,}{{Toloba}
  et~al.}{2023}]{Toloba2023}
{Toloba} E.,  et~al., 2023, \mn@doi [arXiv e-prints]
  {10.48550/arXiv.2305.06369}, \href
  {https://ui.adsabs.harvard.edu/abs/2023arXiv230506369T} {p. arXiv:2305.06369}

\bibitem[\protect\citeauthoryear{{Tremmel}, {Wright}, {Brooks}, {Munshi},
  {Nagai}  \& {Quinn}}{{Tremmel} et~al.}{2020}]{Tremmel2020}
{Tremmel} M.,  {Wright} A.~C.,  {Brooks} A.~M.,  {Munshi} F.,  {Nagai} D.,
  {Quinn} T.~R.,  2020, \mn@doi [\mnras] {10.1093/mnras/staa2015}, \href
  {https://ui.adsabs.harvard.edu/abs/2020MNRAS.497.2786T} {497, 2786}

\bibitem[\protect\citeauthoryear{{Trujillo-Gomez}, {Kruijssen}  \&
  {Reina-Campos}}{{Trujillo-Gomez} et~al.}{2022}]{TrujilloGomez2022}
{Trujillo-Gomez} S.,  {Kruijssen} J.~M.~D.,   {Reina-Campos} M.,  2022, \mn@doi
  [\mnras] {10.1093/mnras/stab3401}, \href
  {https://ui.adsabs.harvard.edu/abs/2022MNRAS.510.3356T} {510, 3356}

\bibitem[\protect\citeauthoryear{{Venhola} et~al.,}{{Venhola}
  et~al.}{2022}]{Venhola2022}
{Venhola} A.,  et~al., 2022, \mn@doi [\aap] {10.1051/0004-6361/202141756},
  \href {https://ui.adsabs.harvard.edu/abs/2022A&A...662A..43V} {662, A43}

\bibitem[\protect\citeauthoryear{{Weinberger} et~al.,}{{Weinberger}
  et~al.}{2017}]{weinberger2017}
{Weinberger} R.,  et~al., 2017, \mn@doi [\mnras] {10.1093/mnras/stw2944}, \href
  {https://ui.adsabs.harvard.edu/abs/2017MNRAS.465.3291W} {465, 3291}

\bibitem[\protect\citeauthoryear{{Weinmann}, {Lisker}, {Guo}, {Meyer}  \&
  {Janz}}{{Weinmann} et~al.}{2011}]{Weinmann2011}
{Weinmann} S.~M.,  {Lisker} T.,  {Guo} Q.,  {Meyer} H.~T.,   {Janz} J.,  2011,
  \mn@doi [\mnras] {10.1111/j.1365-2966.2011.19118.x}, \href
  {https://ui.adsabs.harvard.edu/abs/2011MNRAS.416.1197W} {416, 1197}

\bibitem[\protect\citeauthoryear{{Wolf}, {Martinez}, {Bullock}, {Kaplinghat},
  {Geha}, {Mu{\~n}oz}, {Simon}  \& {Avedo}}{{Wolf} et~al.}{2010}]{Wolf2010}
{Wolf} J.,  {Martinez} G.~D.,  {Bullock} J.~S.,  {Kaplinghat} M.,  {Geha} M.,
  {Mu{\~n}oz} R.~R.,  {Simon} J.~D.,   {Avedo} F.~F.,  2010, \mn@doi [\mnras]
  {10.1111/j.1365-2966.2010.16753.x}, \href
  {https://ui.adsabs.harvard.edu/abs/2010MNRAS.406.1220W} {406, 1220}

\bibitem[\protect\citeauthoryear{{Yagi}, {Koda}, {Komiyama}  \&
  {Yamanoi}}{{Yagi} et~al.}{2016}]{Yagi2016}
{Yagi} M.,  {Koda} J.,  {Komiyama} Y.,   {Yamanoi} H.,  2016, \mn@doi [\apjs]
  {10.3847/0067-0049/225/1/11}, \href
  {https://ui.adsabs.harvard.edu/abs/2016ApJS..225...11Y} {225, 11}

\bibitem[\protect\citeauthoryear{Yahagi \& Bekki}{Yahagi \&
  Bekki}{2005}]{yahagi2005}
Yahagi H.,  Bekki K.,  2005, \mn@doi [Monthly Notices of the Royal Astronomical
  Society: Letters] {10.1111/j.1745-3933.2005.00111.x}, 364, L86

\bibitem[\protect\citeauthoryear{{van Dokkum}, {Abraham}, {Merritt}, {Zhang},
  {Geha}  \& {Conroy}}{{van Dokkum} et~al.}{2015a}]{vanDokkum2015a}
{van Dokkum} P.~G.,  {Abraham} R.,  {Merritt} A.,  {Zhang} J.,  {Geha} M.,
  {Conroy} C.,  2015a, \mn@doi [\apjl] {10.1088/2041-8205/798/2/L45}, \href
  {https://ui.adsabs.harvard.edu/abs/2015ApJ...798L..45V} {798, L45}

\bibitem[\protect\citeauthoryear{{van Dokkum} et~al.,}{{van Dokkum}
  et~al.}{2015b}]{vanDokkum2015b}
{van Dokkum} P.~G.,  et~al., 2015b, \mn@doi [\apjl]
  {10.1088/2041-8205/804/1/L26}, \href
  {https://ui.adsabs.harvard.edu/abs/2015ApJ...804L..26V} {804, L26}

\bibitem[\protect\citeauthoryear{{van Dokkum} et~al.,}{{van Dokkum}
  et~al.}{2016}]{vanDokkum2016}
{van Dokkum} P.,  et~al., 2016, \mn@doi [\apjl] {10.3847/2041-8205/828/1/L6},
  \href {http://adsabs.harvard.edu/abs/2016ApJ...828L...6V} {828, L6}

\bibitem[\protect\citeauthoryear{{van Dokkum} et~al.,}{{van Dokkum}
  et~al.}{2017}]{vanDokkum2017}
{van Dokkum} P.,  et~al., 2017, \mn@doi [\apjl] {10.3847/2041-8213/aa7ca2},
  \href {https://ui.adsabs.harvard.edu/abs/2017ApJ...844L..11V} {844, L11}

\bibitem[\protect\citeauthoryear{{van Dokkum} et~al.,}{{van Dokkum}
  et~al.}{2018}]{vanDokkum2018}
{van Dokkum} P.,  et~al., 2018, \mn@doi [\apjl] {10.3847/2041-8213/aab60b},
  \href {https://ui.adsabs.harvard.edu/abs/2018ApJ...856L..30V} {856, L30}

\bibitem[\protect\citeauthoryear{{van Dokkum} et~al.,}{{van Dokkum}
  et~al.}{2019}]{vanDokkum2019}
{van Dokkum} P.,  et~al., 2019, \mn@doi [\apj] {10.3847/1538-4357/ab2914},
  \href {https://ui.adsabs.harvard.edu/abs/2019ApJ...880...91V} {880, 91}

\bibitem[\protect\citeauthoryear{{van Dokkum} et~al.,}{{van Dokkum}
  et~al.}{2022}]{vanDokkum2022}
{van Dokkum} P.,  et~al., 2022, arXiv e-prints, \href
  {https://ui.adsabs.harvard.edu/abs/2022arXiv220707129V} {p. arXiv:2207.07129}

\bibitem[\protect\citeauthoryear{{van der Burg, Remco F. J.}, {Muzzin, Adam}
  \& {Hoekstra, Henk}}{{van der Burg, Remco F. J.}
  et~al.}{2016}]{vanderBurg2016}
{van der Burg, Remco F. J.} {Muzzin, Adam}  {Hoekstra, Henk} 2016, \mn@doi
  [A\&A] {10.1051/0004-6361/201628222}, 590, A20

\bibitem[\protect\citeauthoryear{{van der Burg} et~al.,}{{van der Burg}
  et~al.}{2017}]{vanderBurg2017}
{van der Burg} R. F.~J.,  et~al., 2017, \mn@doi [\aap]
  {10.1051/0004-6361/201731335}, \href
  {https://ui.adsabs.harvard.edu/abs/2017A&A...607A..79V} {607, A79}

\bibitem[\protect\citeauthoryear{Łokas \& Mamon}{Łokas \&
  Mamon}{2001}]{lokas2001}
Łokas E.~L.,  Mamon G.~A.,  2001, \mn@doi [Monthly Notices of the Royal
  Astronomical Society] {10.1046/j.1365-8711.2001.04007.x}, 321, 155

\makeatother
\end{thebibliography}






\bsp	
\label{lastpage}
\end{document}